\documentclass[traditabstract,a4]{aa} 
\usepackage{graphicx}
\usepackage{txfonts}
\usepackage{natbib}
\usepackage{ulem}
\def\uss0943{MRC\,0943$-$242}
\def\pks1138{MRC\,1138$-$262}
\begin{document}

\title{Optical and near-IR spectroscopy of candidate red galaxies \\
 in two $z\sim2.5$ proto-clusters\thanks{Based in part on data collected at Subaru Telescope, which
is operated by the National Astronomical Observatory of Japan, and in part on data collected with ESO's Very Large Telescope UT1/Antu, under program ID 080.A-0463(B). and ESO's New Technology Telescope under program ID 076.A-0670(B) }} 

\subtitle{}

\author{M. Doherty\inst{1}
          \and
          M. Tanaka\inst{2}
          \and
          C. De Breuck\inst{2}
          \and
          C. Ly\inst{3}
          \and 
          T. Kodama\inst{4}
          \and
	  J. Kurk\inst{5,6}
          \and
          N. Seymour\inst{7}
          \and
          J. Vernet\inst{2}
	  \and
          D. Stern\inst{8}
          \and
          B. Venemans\inst{2}
	  \and
	  M. Kajisawa\inst{4,9}
          \and
          I. Tanaka\inst{10}
}

\institute{European Southern Observatory, ESO Santiago, Alonso
de Cordova 3107, Vitacura, Santiago; \email{mdoherty@eso.org}
         \and
	 European Southern Observatory, ESO Garching, Karl-Schwarzschild-Str 2, D-85748, Garching, Germany
         \and
	 Department of Physics and Astronomy, University of California, Los Angeles, Box 951547, CA 90095, USA
	 \and
         National Astronomical Observatory of Japan, Mitaka, Tokyo 181-8588, Japan 
         \and
	 Max Planck Institut f\"{u}r Astronomie, K\"{o}nigstuhl 17, D-69117, Heidelberg, Germany 
	 \and
	 Max Planck Institut f\"{u}r Extraterrestrische Physik, Giessenbachstrasse, D-68165, Garching, Germany 
	 \and         Mullard Space Science Laboratory, UCL, Holmbury St Mary, Dorking, Surrey RH5 6NT, UK
	 \and
         Jet Propulsion Laboratory, California Institute of Technology, 4800 Oak Grove Dr., Pasadena, CA91109, USA
         \and
          Subaru Telescope, National Astronomical Observatory of Japan, 650 North Aohoku Place, Hilo, HI96720, USA 
	 \and
	 Astronomical Institute, Tohoku University, Aramaki, Aoba, Sendai 980-8578, Japan 
}

\date{Received July 10, 2009; accepted November 3, 2009}

\abstract{
  We present a spectroscopic campaign to follow-up red colour-selected
  candidate massive galaxies in two high redshift proto-clusters
  surrounding radio galaxies.  We observed a total of 57 galaxies in
  the field of \uss0943\ ($z=2.93$) and 33 in the field of \pks1138\
  ($z=2.16$) with a mix of optical and near-infrared multi-object
  spectroscopy.

  We confirm two red galaxies in the field of \pks1138\ at the
  redshift of the radio galaxy. Based on an analysis of their
  spectral energy distributions, and their derived star formation
  rates from the H$\alpha$ and 24\,$\mu$m flux, one object belongs to
  the class of dust-obscured star-forming red galaxies, while the
  other is evolved with little ongoing star formation.  This result
  represents the first red and mainly passively evolving galaxy to be
  confirmed as companion galaxies in a $z>2$ proto-cluster. Both red
  galaxies in \pks1138\ are massive, of the order of $4-6 \times
  10^{11}$ M$_{\odot}$. They lie along a Colour-Magnitude relation
  which implies that they formed the bulk of their stellar population
  around $z=4$.

  In the \uss0943 field we find no red galaxies at the redshift of the
  radio galaxy but we do confirm the effectiveness of our $JHK_s$
  selection of galaxies at $2.3<z<3.1$, finding that 10 out of 18
  (56\%) of $JHK_s$-selected galaxies whose redshifts could be
  measured fall within this redshift range.  We also serendipitously
  identify an interesting foreground structure of 6 galaxies at
  $z=2.6$ in the field of \uss0943.  This may be a proto-cluster
  itself, but complicates any interpretation of the red sequence
  build-up in \uss0943\ until more redshifts can be measured.

  }


\keywords{Galaxies: clusters: individual: \pks1138\ -- Galaxies:
clusters: individual: \uss0943\ -- Galaxies: evolution --
Galaxies: high-redshift -- (Cosmology:) large-scale structure of
Universe -- Infrared: galaxies }

\titlerunning{Spectroscopy of red galaxies in $z\sim2.5$ proto-clusters}

\maketitle
 

\section{Introduction}

At every epoch, galaxies with the reddest colours are known to trace
the most massive objects \citep{tka+05}.  The study of massive galaxies at high
redshift therefore places important constraints on the mechanisms
and physics of galaxy formation \citep[e.g.,][]{hok+09}.  Studies of clusters at both low
and high redshift also yield valuable insights into the assembly
and evolution of large-scale structure, which itself is a sensitive
cosmological probe \citep[e.g.,][]{ecf+96}.

The most distant spectroscopically confirmed cluster lies at $z=1.45$
and has 17 confirmed members, a large fraction of which are consistent
with massive, passively evolving ellipticals \citep{srs+06, hcs+07}.
High redshift proto-clusters have been inferred  through
narrow-band imaging searches for Ly-$\alpha $ emitters in the
vicinity of radio galaxies \citep[e.g.,][]{kpo+04, vrm+05}.  Such
searches have been highly successful, with follow-up spectroscopy
confirming $15-35$ Ly-$\alpha$ emitters per proto-cluster.  However,
the Ly-$\alpha$ emitters are small, faint, blue star-forming galaxies,
and likely constitute a small fraction of both the number of cluster
galaxies and the total mass budget in the cluster.

Massive elliptical galaxies are seen to dominate cluster cores up
to a redshift of $z\approx1$ \citep[e.g.,][]{vsh+01}, with the red
sequence firmly in place by that epoch \citep{ejd+06, bfp+03} and
in fact showing very little evolution even out to $z\sim1.4$
\citep[e.g.,][]{lts+08}.  What remains uncertain is whether this
holds true at yet higher redshift.  The fundamental questions are (1)
at what epoch did massive, rich clusters first begin to assemble,
and (2) from what point do they undergo mostly passive evolution?
There is certainly evidence that the constituents of high redshift
clusters are very different from that of rich clusters in the local
Universe, with the distant clusters containing a higher fraction
of both blue spirals \citep[e.g.,][]{ejd+06} and active galaxies
\citep{gse+09}.  \citet{vs03} found early-type galaxies with
signatures of recent star formation in a $z=1.27$ cluster, perhaps
symptomatic of a recently formed (or still forming) cluster.  Locating
massive galaxies in higher redshift clusters will therefore yield
valuable insight into the earliest stages of cluster formation and
the build-up of complex structures which are observed in the local
Universe.

In two previous papers, we found over-densities of red colour-selected
galaxies in several proto-clusters associated with high redshift radio
galaxies \citep{kkt+06, kkk+07}. Follow up studies of distant red
  galaxies (DRGs) have shown that this colour selection criterion
  contains two populations: (1) dusty star-forming galaxies and (2)
  passively evolved galaxies. The exact fraction between these two
  populations is still a matter of debate
  \citep{for04,gra06,gra07}. With our new $JHK$ colour-selection, we
  hope to be more efficient in identifying the evolved galaxy
  population. We have now embarked upon a spectroscopic follow-up
  investigation with the aim of confirming cluster members and
  determining their masses and recent star formation histories. Such a
  goal is ambitious as redshifts of these red galaxies are generally
  very difficult to obtain, especially amongst the population with no
  on-going star formation (i.e., lacking emission lines).  Here we
present the first results from optical and near-infrared spectroscopy
of targets in two of the best studied proto-clusters, \pks1138\
($z=2.16$) and \uss0943\ ($z=2.93$).

In Section 2 we explain the target selection and in Section 3 outline
the observations and data reduction steps. Section 4 presents the
redshifts identified and in Section 5 we briefly analyse the
properties of the two galaxies discovered at the redshift of \pks1138,
including their relative spatial location to the RG, ages, masses and
star formation rates from spectral energy distribution fits and star
formation rates from the H$\alpha$ emission line flux.  In Section 6
we summarise the results and draw some wider conclusions.

Throughout this paper we assume a cosmology of H$_0$=71~km~s$^{-1}$,
$\Omega_{\rm M}$ = 0.27, $\Omega_{\Lambda}$=0.73 \citep{svp+03} and
magnitudes are on the Vega system.

\section{Target selection}
\subsection{Proto-clusters}
We have concentrated the initial stages of our spectroscopic campaign
on \pks1138\ and \uss0943\ for several reasons.  First, these two
proto-clusters fields were both observed in the space infrared by a
{\it Spitzer} survey of 70 high redshift radio galaxies
\citep[RGs;][]{ssd+07}.  They are also amongst the eight $z>2$ radio
galaxies observed in the broad and narrow-band imaging survey of
\citet{vrm+07}.  Both targets therefore are among the best studied
high-redshift radio galaxies and have a large quantity of broadband
data available from the $U$-band to the {\it Spitzer} bands, over a
wide field of view.  Their redshifts span the $2<z<3$ region in which
it has been shown that the red sequence most likely starts to build up
\citet{kkk+07}, and they therefore probe an interesting and poorly
understood redshift regime for understanding the build-up of red,
massive galaxies in proto-clusters.

Our group previously obtained deep, wide-field near-infrared imaging
of \pks1138\ and \uss0943\ with the Multi-Object InfraRed Camera and
Spectrograph \citep[MOIRCS;][]{sti+08,ist+06}, the relatively new
$4\arcmin \times 7\arcmin$ imager on Subaru \citep{kkk+07}.  The data
were obtained in exceptional conditions, with seeing ranging from
0\farcs5 to 0\farcs7. We use our previously published photometric
  catalogues, and quote the total magnitudes (MAGAUTO in Sextractor)
  for individual band photometry, and 1.5\arcsec\ diameter aperture
  magnitudes for all colours. We refer to \citep{kkk+07} for further
  details on the photometry. We also used an $H-$band image of
  \pks1138\ obtained with the Son of ISAAC (SOFI) on the New
  Technology Telescope (NTT) on UT 2006 March 23. The total exposure
  time was 9720\,s. We reduced the data using the standard procedures
  in IRAF.

Using this data set, we selected galaxies expected to lie at the same
redshifts as the radio galaxies on the basis of their $J - K_s$ or
$JHK_s$ colours.  These deep data indicate a clear excess of the
near-infrared-selected galaxies clustered towards the radio galaxies
\citep[e.g., Figures 3--5 in][]{kkk+07}.  The excess is a factor of
$\approx 2-3$ relative to the GOODS-S field.

\subsection{MRC~1138--262 at $z=2.16$}
The red galaxies in \pks1138\ were selected according to the `classic'
DRG criterion of $J-Ks>2.3$ \citep{vff+04}. We note that in
\citet{kkk+07} the zero points of the photometry in this field were
incorrect by 0.25 and 0.30 magnitudes in $J$ and $K_s$ bands,
respectively, with $J^\prime = J - 0.25$, $K_s^\prime = K_s - 0.30$
for this field, where the primed photometry represents the corrected
values.  Since the corrections for both bands are similar, there is
little change in the sample of DRGs identified, which was the primary
thrust of that paper.  We have adjusted the magnitudes and colours
before revising the target selection for the current paper.  We also
include three near-infrared spectroscopic targets in the field of
\pks1138\ from the {\it Hubble Space Telescope} NICMOS imaging
reported by \citet{zsp+08}, two of which fulfill the DRG criterion and
one of which is slightly bluer. We targeted 33 red galaxies in this
field, down to a $K_s$ magnitude of 21.7 .

\subsection{MRC~0943--242 at $z=2.93$}

\begin{figure}[h]
\centering
\begin{tabular}{cc}
\hspace{-0.5cm}\includegraphics[width=0.8\linewidth]{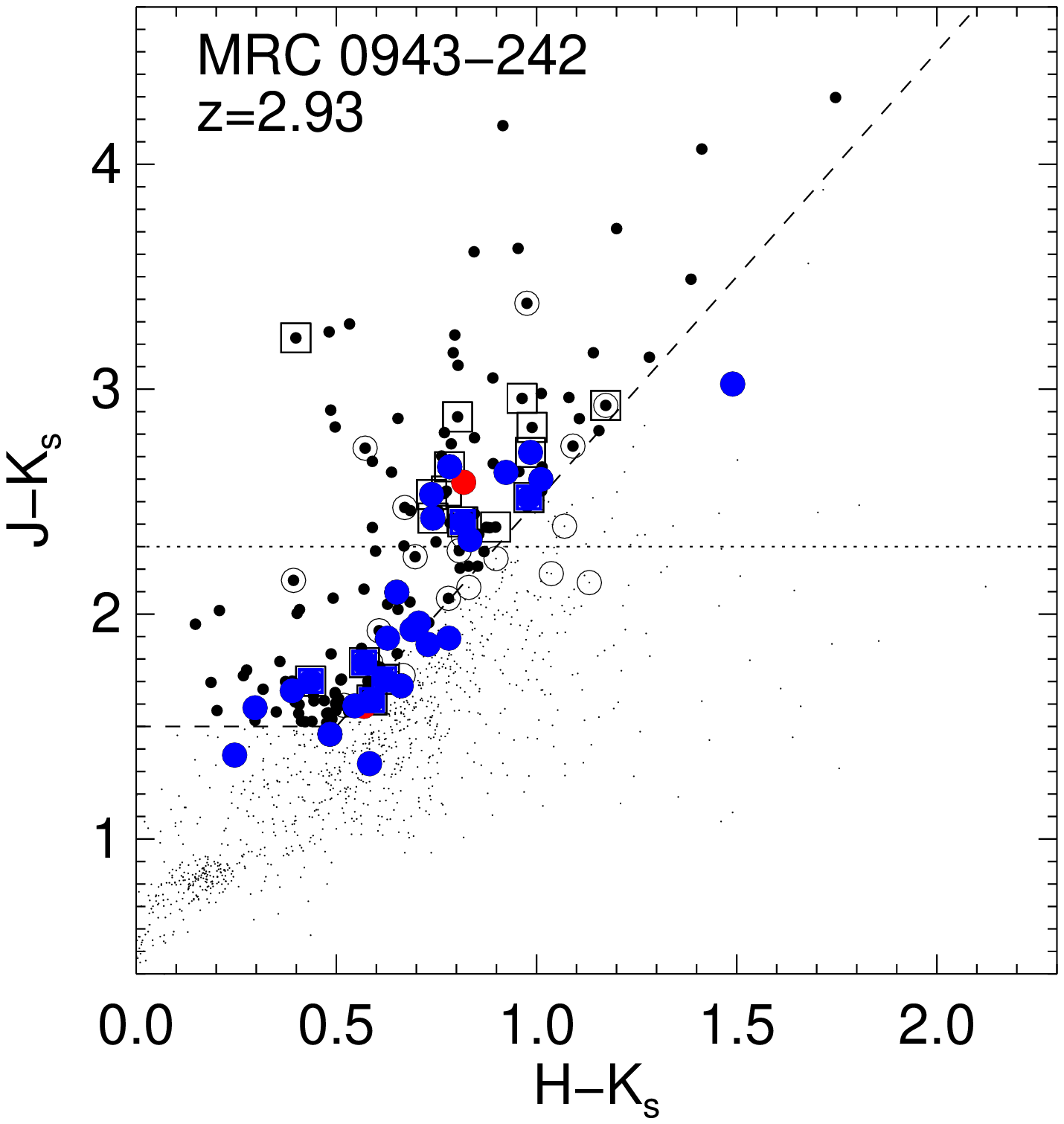} & \hspace{-3cm}\includegraphics[width=0.8\linewidth]{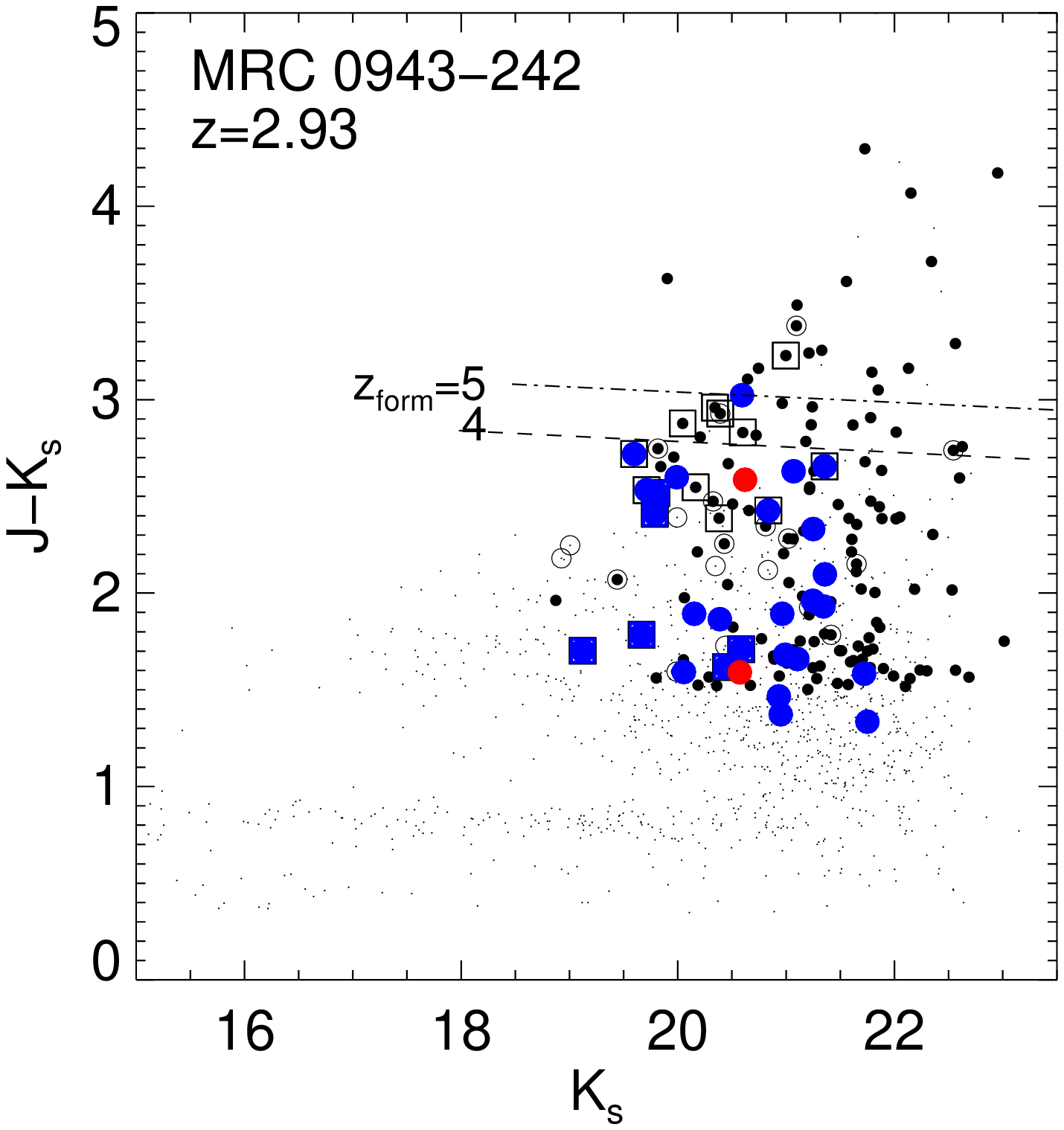}\vspace{-4ex}\\
\hspace{-5.3cm}{\bf (a)} & \hspace{-7.2cm}{\bf (b)} \\
\hspace{-0.5cm}\includegraphics[width=0.8\linewidth]{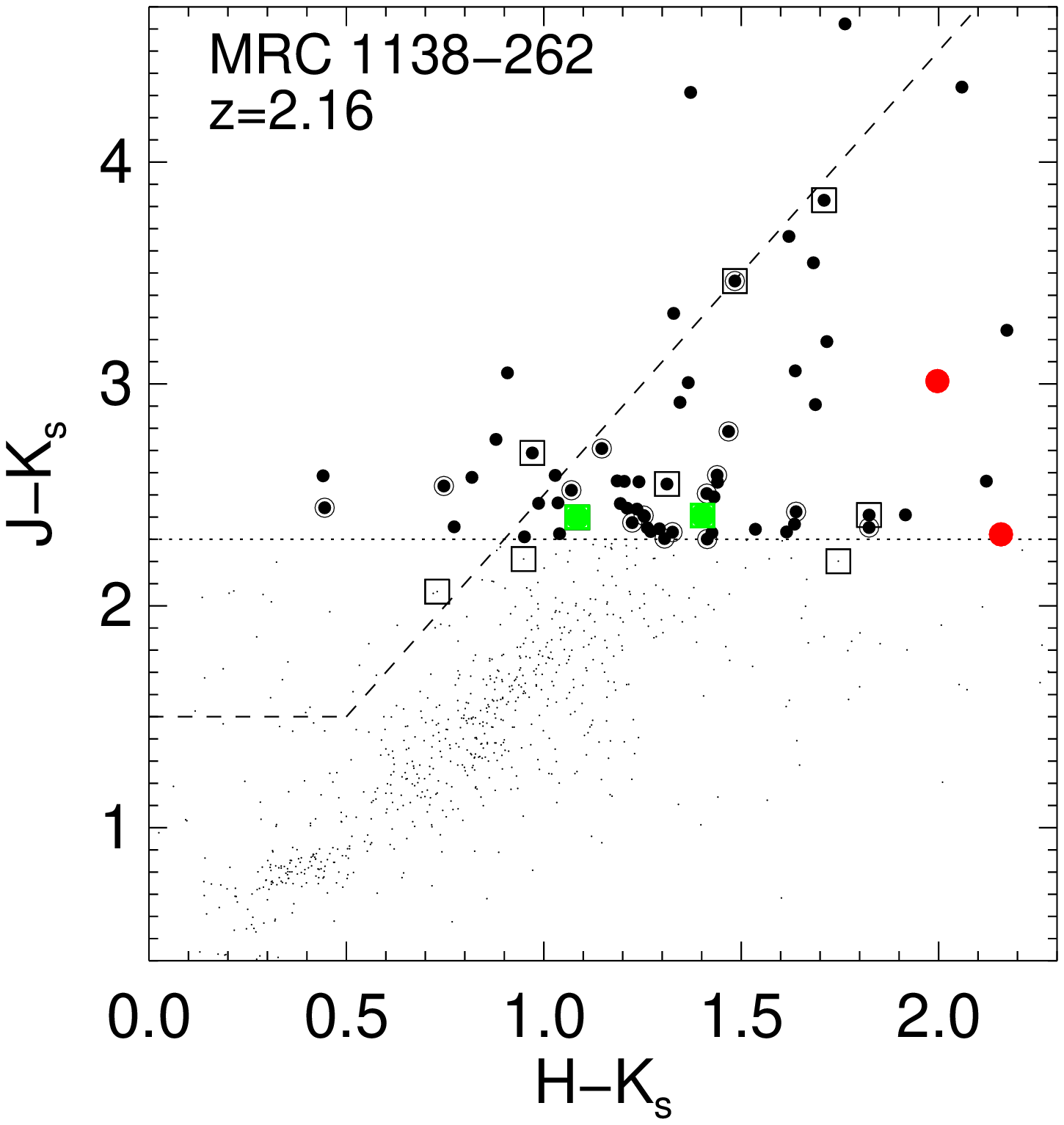} & \hspace{-3cm}\includegraphics[width=0.8\linewidth]{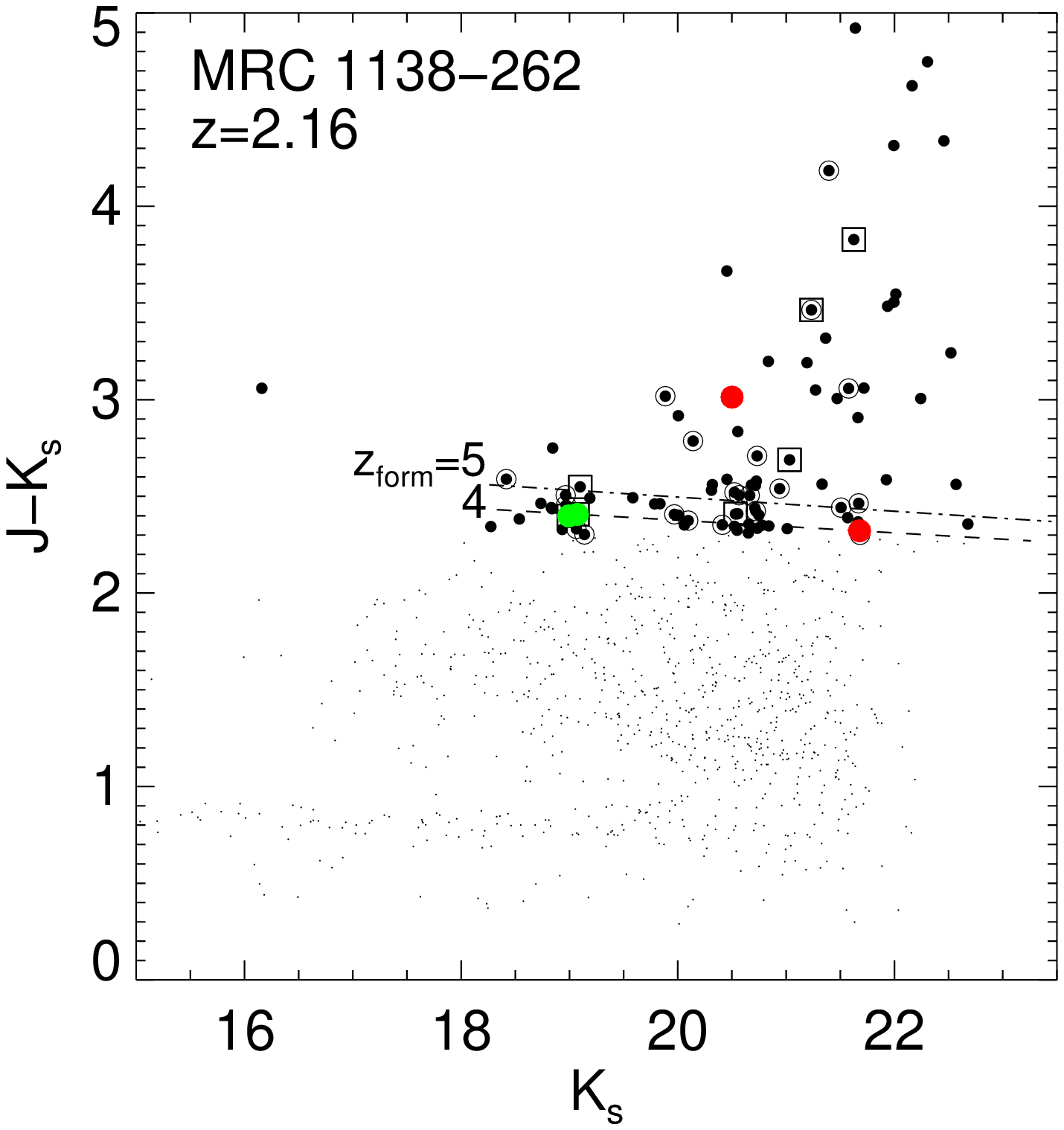}\vspace{-4ex}\\
\hspace{-5.3cm}{\bf (c)} & \hspace{-7.2cm}{\bf (d)} \\
\end{tabular}
\caption{Spectroscopic targets shown in colour-colour (left) and
colour-magnitude (right) space for \uss0943\ (top) and \pks1138\
(bottom).  Objects observed with optical spectroscopy by either
FORS2 or FOCAS are indicated by large circles; objects observed with
near-infrared spectroscopy by MOIRCS are indicated by squares.  With
respect to the radio galaxy, filled blue symbols are confirmed
foreground objects and red filled symbols are background galaxies.
Green symbols show galaxies confirmed to lie at the redshift of the
radio galaxy.}
\label{fig:targ_sel}
\end{figure}

For the redshift $z\sim3$ proto-cluster \uss0943, we define two
classes of colour-selected objects: blue $JHK_s$ galaxies (bJHKs)
defined to have: \\
$J-K_s> 2(H-K_s)+0.5$ and $J-K_s>1.5$, \\
and red $JHK_s$ galaxies (rJHKs) defined to have: \\
$J-K_s> 2(H-K_s)+0.5$ and $J-K_s>2.3$. \\
Figure~\ref{fig:targ_sel} shows these colour cuts and the objects
which were targeted for spectroscopy. Note that although we have not
used $H-$band data in the selection of targets for \pks1138, we
include a $JHK_s$ diagram for that field as a comparison to \uss0943\
- it is obvious from these diagrams that the two colour cuts select
distinct populations of objects.

\begin{figure}[h]
\centering
\includegraphics[width=\linewidth]{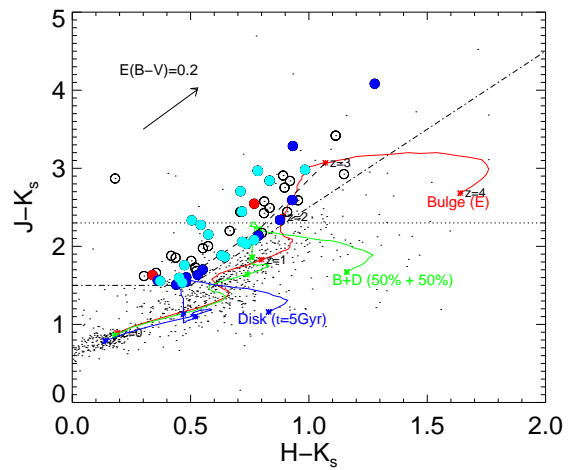}
\caption{Colour selection technique designed to select galaxies at
$2.3<z<3.1$. The solid curves represent the evolutionary tracks of
galaxies over $0<z<4$ with different star formation histories
(passive, intermediate, and active, from top to bottom, respectively)
formed at $z_{\rm form}=5$ \citep{kbb99}. The dashed line connects the
model points at $z=3$.  Objects targeted spectroscopically are
represented by circles; filled circles indicate where a redshift
identification has been made.  Blue circles indicate objects at
$z<2.3$ and red circles indicate sources at $z>3.1$ --- i.e., outside
of the targeted redshift range.  Cyan circles show objects whose
redshifts fall within the targeted range.}
\label{fig:coloursuccess}
\end{figure}

Figure~\ref{fig:coloursuccess} demonstrates how our $JHK_s$ selection
works to select candidate proto-cluster members around \uss0943. The
solid curves represent the evolutionary tracks of galaxies over
$0<z<4$ with different star formation histories (passive,
intermediate, and active) formed at $z_{\rm form}=5$ \citep{kbb99}.
The dashed line connects the model points at $z=3$, and indicates the
region where we expect to find galaxies at a similar redshift to the radio
galaxy.  We therefore apply the colour-cut (shown by the dot-dashed
lines) to exclusively search for galaxies associated with the radio
galaxies at $2.3 < z < 3.1$.  This technique was originally used by
\citet{kkt+06} and has a great advantage over the DRG selection used
at lower redshift since we also select the bluer populations within
the redshift range of interest. We targeted 38 $JHK_s$ selected
galaxies (including the RG itself) down to a magnitude $K_s=22$. In
section 4.1, we will show that we could measure redshifts of half of
these sources with a success rate of our selection technique of 
$\sim$55\%. In the remaining spectroscopic slits, we included 22
additional ``filler'' targets (see Table~\ref{tab:redshifts0943}),
mostly those with colours just outside our selection criterion, but
also two Distant Red Galaxies ($J-K>2.3$) and three Ly$\alpha$
emitters.

Where two or more objects clashed in position, we prioritised objects
with 24$\mu$m detections from the Multiband Imaging Photometer for
{\it Spitzer} \citep[MIPS;][]{rye+04}.  For the optical spectroscopy,
we also prioritised objects with brighter $I$-band magnitudes. After
selecting primary targets according to the above colour criteria, a
couple of additional `filler' targets were selected depending on the
geometry of the mask. These filler objects were drawn from a list of
candidate Ly-$\alpha$ emitters which do not yet have confirmed
redshifts \citep{vrm+07}. Figure~\ref{fig:targ_sel} shows the
distribution of selected targets in colour-colour and colour-magnitude
space.  Whilst by no means complete sample, we believe our sample to
be representative of the candidate objects.

\section{Observations and data reduction}

We obtained optical multi-object spectra of targeted galaxies in the
two proto-clusters using the Faint Object Camera and Spectrograph
\citep[FOCAS;][]{kii+00} on Subaru, and the FOcal Reducer and low
dispersion Spectrograph-2 \citep[FORS2;][]{ar92} on the
VLT\footnote{ESO program 080.A-0463(B).}.  We obtained near-infrared
multi-object spectra of targeted galaxies using MOIRCS on Subaru.  For
one galaxy, we also obtained a spectrum with X-SHOOTER \citep{ddm+06}
on the VLT during a commissioning run.  A summary of the observations
is presented in Table~\ref{tab:obs}.

\subsection{Optical spectra}

We were scheduled two nights, UT 2007 March 14-15, of FOCAS
spectroscopy on Subaru.  However, 1.5 out of the two nights were lost
to bad weather.  In the first half of the second night, we obtained
three hours integration on \uss0943\ in variable seeing.  We used the
300B grism with the L600 filter and binned the readout $3\times2$
(spatial $\times$ spectral).  This set-up gives a dispersion of 1.34
\AA\ pix$^{-1}$ and a spectral resolution of 9 \AA\ full width at half
maximum (FWHM; as measured from sky lines).  There were 22 slits on
the mask, but the spectra for three objects fell on bad CCD defects so
we treat these as unobserved. Two Ly-$\alpha$ emitter candidates were
also placed on the mask to fill in space where there were no suitable
near-infrared selected galaxies.  The data were reduced in a standard
manner using IRAF and custom software. Although the night was not
photometric, spectra were flux calibrated with standard stars observed
on the same night, as it is important to calibrate at least the
relative spectral response given that we are interested in breaks
across the spectra. 

We obtained FORS2/MXU spectra with UT2/Kueyen on UT 2008 March
11-12. The nights were clear with variable seeing averaging around
1\arcsec. The 300V grating was used, providing a dispersion of 3.36
\AA\ pix$^{-1}$ and a resolution of 12 \AA\ (FWHM, measured from sky
lines).  Slitlets were 1\arcsec\ wide and 6$-$10\arcsec\ long.  We
used 2\arcsec\ nods to shift objects along the slitlets, thereby
creating A-B pairs to improve sky subtraction.  We first created
time-contiguous pairs of subtracted frames, and then averaged all of
the A$-$B pairs together in order to improve the resulting sky
subtraction and to reject cosmic rays (in both the positive and
negative images).  These combined frames were then processed using the
FORS2 pipeline (ver. 3.1.7) with standard inputs.  The extracted
one-dimensional spectra were boxcar extracted over a 1\arcsec\
aperture and calibrated with flux standards taken on the same night.
In total, we obtained 10 hours of integration on \uss0943\ and five
hours of integration on \pks1138.

\subsection{Near-infrared spectra}

We obtained MOIRCS spectra for \pks1138\ over the latter halves of UT
2008 January 11-12.  Both nights were photometric, and we obtained a
total of five hours of integration.  Since Chip~1 of MOIRCS was the
engineering grade chip at that time, we gained no useful data on that
chip and essentially lost half of the field coverage. This detector
suffered a loss of sensitivity compared to chip 2. More problematic
was a prominent large ring-like structure with significantly high (and
variable) dark noise, as well as several smaller and patchy
structures, which we were unable to calibrate out successfully.  We
obtained $K$-band spectra with the medium resolution (R1300) grating,
yielding a dispersion of 3.88 \AA\ pix$^{-1}$ and a resolution of
26~\AA\ (FWHM, measured from sky lines).  The objects were nodded
along the slitlets and A-B pairs combined to remove the sky
background.  The frames were then flat-fielded and corrected for
optical distortion using the {\tt mscgeocorr} task in IRAF.  They were
then rotated to correct for the fact that the slits are tilted on the
detector.  We extracted spectra using a box-car summation over the
width of the continuum, or, in the case of no continuum, over the
spatial extent of any detected line emission.

We obtained MOIRCS data for \uss0943\ on UT 2009 January 10, with
four hours total integration time over the latter half of the night.
The low resolution HK500 grating was used, providing a dispersion
of 7.72 \AA\ pix$^{-1}$ and a resolution of 42 \AA\ (FWHM, measured
from sky lines).

\begin{table*}
\caption{Summary of observations.}             
\label{tab:obs}      
\centering                          
\begin{tabular}{r r r r r r r r r r}        
\hline\hline \\
Proto-cluster  & Instrument & FOV & Filter & Grating & Spectral Coverage$^a$ ($\mu$m)& UT Date Obs. & Total Exp. \\
\hline \\
\pks1138 &  MOIRCS & $4\arcmin \times 3\farcm5$ & $K$ & R1300 & $2.0-2.4$ & 11-12 Jan 2008 & 5 hr \\
\pks1138 &  FORS2 & $7\arcmin \times 7\arcmin$ & -- & 300V & $0.33-0.9$ & 11-12 Mar 2008 & 5 hr \\
\uss0943 &  FORS2 & $7\arcmin \times 7\arcmin$ & -- & 300V & $0.33-0.9$ & 11-12 Mar 2008 & 10 hr \\
\uss0943 &  FOCAS & $6\arcmin $ diameter & L600 & 300B & $0.37-0.6$ & 15 Mar 2007 & 3 hr \\
\uss0943 &  MOIRCS & $4\arcmin \times 7\arcmin$ & OC\_HK & HK500 & $1.3-2.5$ & 10 Jan 2009 & 4 hr \\
\uss0943 &  X-SHOOTER & $1\arcsec \times 11\arcsec$ &    --  &    -- & $0.3-2.5$ &  5 Jun 2009 & 1 hr \\
\hline \\
\multicolumn{7}{l}{$^a$ Exact spectral coverage for each object depends on the position of the slit on the detector.}
\end{tabular}
\end{table*} 

\subsection{UV/Optical/NIR spectrum}

We also obtained data for a single target in the field of
\uss0943\ with X-SHOOTER during a commissioning run on 2009 June
5 under variable seeing conditions (between 0.5\arcsec\ and
2\arcsec\ during the exposure).  Slit widths of 1.0\arcsec,
0.9\arcsec\ and 0.9\arcsec\ were used in the UV--blue, visual--red and
near--IR arm, respectively, resulting in resolutions of R$_{\rm
  UVB}$=5100, R$_{\rm VIS}$=8800 and R$_{\rm NIR}$=5100.  The data were
reduced using a beta version of the pipeline developed by the
X-SHOOTER consortium \citep{grf+06}. The visual-red arm spectrum
(550nm--1000nm) presented below is a combination of three 1200s
exposures. The near-IR arm spectrum (1000nm-2500nm) is a combination
of two 1300s exposures in a nodding sequence.

\section{Results/redshift identifications}

Redshifts were identified by visual inspection of both the processed,
stacked two-dimensional data and the extracted, one-dimensional
spectra.  We based redshifts on both emission and absorption line
features, as well as possible continuum breaks and assigned confidence flags.

Table~\ref{tab:redshifts0943} shows the sources observed with optical
spectroscopy and the corresponding redshift identifications and
quality flags.  Quality 1 indicates confirmed, confident redshifts;
quality 2 indicates doubtful redshifts; quality 3 indicates that we
detected continuum emission but no identification was made, typically
due to the low signal-noise ratio (S/N) of the data; and quality 4
indicated that no emission was detected.  In some cases we can infer
an upper limit to the redshift if the continuum extends to the blue
wavelength cutoff of the optical spectrum.  This is useful for
distinguishing foreground objects. In practice, due to the lower
efficiency of the grisms, the spectra become too noisy to distinguish
the continuum below 3900 \AA\ for FORS and 4100 \AA\ for FOCAS,
corresponding to Ly-$\alpha$ at z=2.3 and z=2.4, respectively.

\subsection{\uss0943}

\begin{table*}
\caption{Results of optical and near-IR spectroscopy in the \uss0943\ field.}  
\label{tab:redshifts0943}      
\centering                          
\begin{tabular}{r r r r r r r r r l r r}        
\hline\hline \\
ID    & Type$^a$ & RA(J2000)    & Dec.(J2000)   & $V_{total}$ &  $I_{total}$    &  $J_{total}$  & $K_{s, total}$ & $J-K_s$ (1\farcs5) & Instr.  & $z$      & Quality$^b$ \\
\hline 
897   & bJHK & 09 45 25.34 & $-$24 25 40.3 &$>$26.8&$>$25.6& 23.69 & 21.65 & 2.15 &  FORS2 &   ...   & 4 \\
420   & rJHK & 09 45 25.57 & $-$24 28 59.8 &...$^c$&...$^c$& 23.65 & 21.35 & 2.65 &  FOCAS &   ...   & 3 \\ 
      &      &             &               &       &       &       &       &      & MOIRCS & 2.174   & 2 \\
413   & rJHK & 09 45 25.74 & $-$24 30 28.8 &$>$26.8&$>$25.6& 21.85 & 19.79 & 2.41 & MOIRCS &   ...   & 3 \\
410   & bJHK & 09 45 26.07 & $-$24 30 20.2 & 23.73 & 22.85 & 21.78 & 20.36 & 1.52 & MOIRCS & 1.420   & 2 \\
402   & rJHK & 09 45 26.15 & $-$24 29 38.2 &$>$26.8&$>$25.6& 23.38 & 20.34 & 2.96 & MOIRCS & 2.466   & 2 \\
873   & bJHK & 09 45 26.16 & $-$24 26 48.2 & 24.54 & 23.27 & 21.64 & 19.99 & 1.59 &  FOCAS & 0.272   & 1 \\
860   &filler& 09 45 26.55 & $-$24 28 14.8 & 27.2  & 24.99 & 21.28 & 19.01 & 2.25 &  FORS2 &   ...   & 4 \\
387   & bJHK & 09 45 26.90 & $-$24 31 08.2 &$>$26.8&$>$25.6& 23.29 & 20.02 & 2.28 &  FORS2 &   ...   & 4 \\
845   & rJHK & 09 45 27.07 & $-$24 28 44.5 & 25.3  & 24.01 & 22.24 & 19.60 & 2.72 &  FORS2 & 2.335   & 2 \\
      &      &             &               &       &       &       &       &      & MOIRCS &   ...   & 3 \\
373   &filler& 09 45 27.27 & $-$24 30 37.7 & 24.70 & 23.52 & 21.63 & 20.06 & 1.59 &  FORS2 &  $<$2.3 & 3 \\
830   & rJHK & 09 45 27.39 & $-$24 26 45.9 &$>$26.8&$>$25.6& 22.94 & 20.39 & 2.93 &  FORS2 &   ...   & 4 \\
      &      &             &               &       &       &       &       &      & MOIRCS &   ...   & 4 \\
368   &filler& 09 45 27.46 & $-$24 30 46.8 & 24.68 & 22.79 & 20.95 & 18.93 & 2.18 &  FOCAS &   ...   & 4 \\
803   & bJHK & 09 45 28.25 & $-$24 26 03.8 & 24.31 & 23.82 & 22.89 & 21.10 & 1.66 &  FORS2 & 2.68    & 1 \\
792   & bJHK & 09 45 28.52 & $-$24 26 16.8 & 24.74 & 23.19 & 21.43 & 19.67 & 1.78 &  FORS2 & 1.408   & 1 \\ 
      &      &             &               &       &       &       &       &      &  FOCAS & $<$2.4  & 3 \\
      &      &             &               &       &       &       &       &      & MOIRCS &   ...   & 3 \\
      &      &             &               &       &       &       &       &      & XSHOOT & 1.410   & 1 \\
341   & bJHK & 09 45 28.76 & $-$24 29 10.9 & 25.20 & 23.12 & 21.53 & 19.44 & 2.07 &  FOCAS & $<$2.4  & 3 \\  
775   &filler& 09 45 29.12 & $-$24 28 23.9 & 26.47 & 24.60 & 22.75 & 20.35 & 2.14 &  FORS2 &   ...   & 4 \\
760   & rJHK & 09 45 29.58 & $-$24 28 02.5 &$>$26.8&$>$25.6& 23.70 & 21.25 & 2.33 &  FORS2 & 2.65    & 2 \\
749   & rJHK & 09 45 29.91 & $-$24 27 24.7 &$>$26.8&$>$25.6& 22.01 & 19.71 & 2.53 &  FORS2 & 2.635   & 1 \\ 
      &      &             &               &       &       &       &       &      & MOIRCS & ...     & 3 \\ 
722   &filler& 09 45 30.36 & $-$24 25 26.8 & 23.96 & 22.40 & 22.18 & 20.44 & 1.62 &  FORS2 &   ...   & 3 \\
      &      &             &               &       &       &       &       &      & MOIRCS & 1.507   & 1 \\
700   & bJHK & 09 45 31.14 & $-$24 27 01.3 & 24.58 & 23.91 & 22.75 & 20.43 & 2.26 &  FOCAS & 2.386   & 1 \\
267   & rJHK & 09 45 31.31 & $-$24 29 34.8 & 24.96 & 24.60 & 23.24 & 20.83 & 2.43 &  FORS2 & $<$2.3  & 3 \\
      &      &             &               &       &       &       &       &      &  FOCAS &   ...   & 4 \\  
      &      &             &               &       &       &       &       &      & MOIRCS & 2.419   & 1 \\ 
694   &filler& 09 45 31.38 & $-$24 27 39.1 & 26.81 & 24.43 & 23.01 & 21.21 & 1.93 &  FORS2 &   ...   & 4 \\
690   &filler& 09 45 31.50 & $-$24 27 17.8 & 24.31 & 23.78 & 22.48 & 20.99 & 1.68 &  FOCAS & $<$2.4  & 3 \\
686   & rJHK & 09 45 31.58 & $-$24 26 41.0 &$>$26.8&$>$25.6& 23.14 & 20.60 & 2.83 & MOIRCS &   ...   & 3 \\
250   &filler& 09 45 31.77 & $-$24 29 24.4 &$>$26.8&$>$25.6& 22.34 & 20.59 & 1.71 & MOIRCS &   ...   & 3 \\
675   & rJHK & 09 45 31.99 & $-$24 28 14.8 & 24.96 & 24.51 & 22.89 & 20.33 & 2.47 &  FOCAS & 2.520   & 1 \\
241   & bJHK & 09 45 32.28 & $-$24 28 50.6 &$>$26.8&$>$25.6& 22.88 & 20.96 & 1.89 &  FORS2 & 1.16    & 2 \\
660   & rJHK & 09 45 32.53 & $-$24 26 38.0 & 25.06 & 24.26 & 22.87 & 21.07 & 2.63 &  FORS2 & 1.12    & 2 \\
HzRG  & bJHK & 09 45 32.76 & $-$24 28 49.3 &  ...  &  ...  & 20.65 & 19.12 & 1.70 & MOIRCS & 2.923   & 1 \\
221   & rJHK & 09 45 32.81 & $-$24 29 28.2 &$>$26.8&$>$25.6& 22.26 & 20.38 & 2.39 & MOIRCS &   ...   & 4 \\
646   & bJHK & 09 45 32.97 & $-$24 27 59.6 & 25.73 & 24.26 & 23.35 & 21.34 & 1.93 &  FOCAS &   ...   & 4 \\
623   & rJHK & 09 45 33.66 & $-$24 27 13.0 & 26.45 & 25.04 & 24.85 & 22.54 & 2.74 &  FORS2 &   ...   & 4 \\
175   &filler& 09 45 35.06 & $-$24 29 24.3 & 25.02 & 24.09 & 22.23 & 20.15 & 1.89 &  FORS2 & $<$2.8  & 3 \\
174   & rJHK & 09 45 35.11 & $-$24 31 34.6 & 25.46 & 25.37 & 22.51 & 19.82 & 2.75 &  FOCAS &   ...   & 4 \\  
167   & DRG  & 09 45 35.32 & $-$24 29 46.8 &$>$26.8&$>$25.6& 22.24 & 20.59 & 3.02 &  FORS2 & 0.464   & 2 \\
153   & rJHK & 09 45 35.72 & $-$24 31 38.1 & 26.51 & 23.66 & 23.00 & 20.62 & 2.59 &  FORS2 & 3.94    & 2 \\
143   & bJHK & 09 45 36.10 & $-$24 30 34.4 & 24.98 & 24.00 & 23.33 & 21.25 & 1.96 &  FOCAS &   ...   & 3 \\  
139   & rJHK & 09 45 36.34 & $-$24 29 56.9 &$>$26.8&$>$25.6& 21.99 & 19.80 & 2.52 & MOIRCS &   ...   & 3 \\
138   &filler& 09 45 36.42 & $-$24 30 25.6 & 24.29 & 23.65 & 22.27 & 20.95 & 1.37 &  FORS2 & 1.34    & 1 \\
544   & bJHK & 09 45 36.52 & $-$24 26 11.8 & 24.59 & 23.75 & 23.14 & 21.72 & 1.58 &  FOCAS &   ...   & 3 \\
130   & rJHK & 09 45 36.82 & $-$24 29 51.8 &$>$26.8& 25.33 & 23.53 & 19.90 & 3.63 & MOIRCS &   ...   & 3 \\
123   &filler& 09 45 37.03 & $-$24 31 18.1 &$>$26.8&$>$25.6& 21.60 & 19.78 & 1.93 &  FORS2 &   ...   & 3 \\
116   &filler& 09 45 37.23 & $-$24 31 48.5 & 26.45 & 23.78 & 23.11 & 21.75 & 1.34 &  FORS2 & 2.58    & 2 \\
114   & rJHK & 09 45 37.23 & $-$24 29 59.2 &$>$26.8&$>$25.6& 24.91 & 21.10 & 3.38 &  FORS2 &   ...   & 4 \\
110   &filler& 09 45 37.26 & $-$24 30 20.8 &$>$26.8&$>$25.6& 22.22 & 20.39 & 1.87 &  FORS2 & 2.65    & 1 \\
103   & bJHK & 09 45 37.41 & $-$24 30 12.1 & 24.81 & 24.23 & 23.61 & 21.36 & 2.10 &  FORS2 & 2.62    & 1 \\
      &      &             &               &       &       &       &       &      &  FOCAS & 2.625   & 1 \\
522   & bJHK & 09 45 37.44 & $-$24 27 50.4 & 27.5  & 25.23 & 23.15 & 21.42 & 1.78 &  FORS2 &   ...   & 4 \\
100   & rJHK & 09 45 37.47 & $-$24 29 09.5 & 24.61 & 23.73 & 23.10 & 20.16 & 2.55 & MOIRCS & 2.649   & 2 \\
511   & rJHK & 09 45 37.84 & $-$24 28 43.9 & 25.7  & 24.91 & 22.75 & 20.04 & 2.88 & MOIRCS &   ...   & 3 \\
498   & DRG  & 09 45 38.26 & $-$24 28 08.6 & 26.4  & 24.59 & 22.34 & 19.99 & 2.39 &  FOCAS &   ...   & 4 \\  
 72   & rJHK & 09 45 38.30 & $-$24 30 46.0 &$>$26.8&$>$25.6& 22.86 & 20.81 & 2.35 &  FORS2 &   ...   & 4 \\
495   &filler& 09 45 38.33 & $-$24 27 03.3 &$>$26.8&$>$25.6& 22.20 & 20.44 & 1.73 &  FORS2 &   ...   & 3 \\
 64   &filler& 09 45 38.82 & $-$24 30 39.0 & 26.0  & 25.07 & 22.86 & 20.83 & 2.12 &  FOCAS &   ...   & 4 \\
 39   &filler& 09 45 39.72 & $-$24 32 02.6 & 25.4  & 24.13 & 22.22 & 20.57 & 1.59 &  FORS2 & 3.76    & 2 \\
 34   & rJHK & 09 45 39.92 & $-$24 29 19.7 & 27.9  & 25.75 & 22.60 & 19.99 & 2.60 & MOIRCS &   ...   & 3 \\
 23   & rJHK & 09 45 40.40 & $-$24 29 00.6 &$>$26.8&$>$25.6& 24.31 & 21.00 & 3.23 & MOIRCS &   ...   & 4 \\
 16   &filler& 09 45 40.57 & $-$24 31 24.7 & 24.16 & 23.48 & 22.34 & 20.93 & 1.47 &  FORS2 & 2.44    & 1 \\
      &      &             &               &       &       &       &       &      &  FOCAS & $<$2.7  & 2 \\
\hline
LAE1141 & LAE  & 09 45 29.518 & $-$24 29 15.72 & \multicolumn{5}{c}{Ly-$\alpha$ emitter candidate} & MOIRCS & 3.495   & 2 \\ 
LAE1618 & LAE  & 09 45 29.582 & $-$24 28 02.45 & \multicolumn{5}{c}{Ly-$\alpha$ emitter candidate} & FOCAS  &   ...   & 4 \\ 
LAE381  & LAE  & 09 45 37.759 & $-$24 31 10.82 & \multicolumn{5}{c}{Ly-$\alpha$ emitter candidate} & FOCAS  & 2.935   & 1 \\ 
\hline
\multicolumn{11}{l}{$^a$ rJHK, bJHK: see text; DRG: Distant Red Galaxy, $J-K>2.3$.} \\ 
\multicolumn{11}{l}{$^b$ See text for definition of quality flags.} \\
\multicolumn{11}{l}{$^c$ Photometry contaminated by a blue $z$=0.458 foreground galaxy.} \\

\end{tabular}
\end{table*}

\begin{table*}
\caption{Results of optical and near-IR spectroscopy in the \pks1138\ field.}
\label{tab:redshifts1138}      		
\centering                          	
\begin{tabular}{r r r r r r r r l r r}  
\hline\hline \\
ID    &  RA(J2000)    & Dec.(J2000)   & $B_{total}$ &  $I_{total}$    &  $J_{total}$  & $K_{s, total}$ & $J-K_s$ (1\farcs5) & Instr.  & $z$      & Quality$^a$ \\
\hline \\
493   & 11 40 33.42 & $-$26 29 14.6 & $>$27 & $>$26.2 &   22.89 &  20.55 &  2.56 &  FORS2 &   ...   & 4 \\ 
867   & 11 40 34.09 & $-$26 30 38.7 & $>$27 & 25.32   & $>$23.3 &  21.30 &  2.49 &  FORS2 &   ...   & 4 \\ 
156   & 11 40 34.82 & $-$26 27 46.4 & $>$27 & $>$26.2 & $>$23.3 &  21.15 &  4.23 &  FORS2 &   ...   & 4 \\ 
174   & 11 40 35.45 & $-$26 27 51.8 & 25.35 & 25.06   & $>$23.3 & $>$22  &  2.51 &  FORS2 &   ...   & 3 \\ 
676   & 11 40 36.96 & $-$26 29 55.6 & $>$27 & $>$26.2 &   22.57 &  19.79 &  3.07 &  FORS2 &   ...   & 4 \\ 
808   & 11 40 38.19 & $-$26 30 24.3 & $>$27 & 24.87   &   22.91 &  20.44 &  2.56 &  FORS2 &   ...   & 4 \\ 
552   & 11 40 38.28 & $-$26 29 25.1 & $>$27 & 25.37   &   22.67 &  20.47 &  2.41 & MOIRCS &   ...   & 4 \\
364   & 11 40 39.76 & $-$26 28 45.4 & 24.17 & 22.87   &   21.24 &  19.03 &  2.21 & MOIRCS &   ...   & 3 \\
877   & 11 40 40.24 & $-$26 30 41.2 & 25.96 & 24.91   &   23.44 &  21.30 &  2.35 &  FORS2 &   ...   & 3 \\ 
341   & 11 40 41.47 & $-$26 28 37.2 & $>$27 & 24.81   &   23.63 &  20.48 &  3.06 &  FORS2 & 3.263   & 1 \\ 
147   & 11 40 43.48 & $-$26 27 44.1 & $>$27 & $>$26.2 &   21.98 &  20.04 &  2.43 &  FORS2 &   ...   & 3 \\ 
369   & 11 40 43.46 & $-$26 28 45.2 & $>$27 & $>$26.2 & $>$23.3 &  21.30 &  3.83 & MOIRCS &   ...   & 4 \\ 
905   & 11 40 44.01 & $-$26 30 47.0 & $>$27 & $>$26.2 &   23.45 &  20.86 &  2.69 & MOIRCS &   ...   & 4 \\
456   & 11 40 44.28 & $-$26 29 07.7 & $>$27 & 24.71   &   21.28 &  18.93 &  2.45 &  FORS2 &   ...   & 3 \\ 
      &             &               &       &         &         &        &       & MOIRCS & 2.172   & 2 \\ 
464   & 11 40 46.09 & $-$26 29 11.5 & $>$27 & $>$26.2 &   21.54 &  19.06 &  2.41 & MOIRCS & 2.149   & 2 \\ 
558   & 11 40 46.52 & $-$26 29 27.1 & $>$27 & 24.71   &   21.64 &  19.03 &  2.55 & MOIRCS &   ...   & 4 \\ 
476   & 11 40 46.68 & $-$26 29 10.4 & $>$27 & 24.90   &   22.30 &  19.86 &  2.46 &  FORS2 &   ...   & 3 \\ 
605   & 11 40 47.58 & $-$26 29 38.2 & $>$27 & $>$26.2 & $>$23.3 &  21.07 &  3.51 &  FORS2 &   ...   & 4 \\ 
      &             &               &       &         &         &        &       & MOIRCS &   ...   & 4 \\ 
830   & 11 40 48.36 & $-$26 30 30.6 & $>$27 & 20.48   &   20.67 &  18.86 &  2.20 & MOIRCS &   ...   & 3 \\
492   & 11 40 49.59 & $-$26 29 07.7 & $>$27 & 25.60   &   23.09 &  20.02 &  2.84 &  FORS2 &   ...   & 4 \\ 
392   & 11 40 50.35 & $-$26 28 49.6 & $>$27 & $>$26.2 &   23.47 &  20.56 &  2.47 &  FORS2 &   ...   & 4 \\ 
573   & 11 40 50.75 & $-$26 29 32.5 & $>$27 & 24.68   &   21.98 &  20.04 &  2.06 & MOIRCS &   ...   & 4 \\ 
601   & 11 40 51.30 & $-$26 29 38.5 & $>$27 & 23.45   &   21.27 &  19.05 &  2.35 &  FORS2 &   ...   & 3 \\ 
506   & 11 40 53.13 & $-$26 29 18.1 & $>$27 & 23.55   &   21.33 &  19.01 &  2.38 &  FORS2 &   ...   & 3 \\ 
165   & 11 40 54.01 & $-$26 27 48.1 & $>$27 & $>$26.2 &   23.35 &  20.85 &  2.59 &  FORS2 &   ...   & 4 \\ 
241   & 11 40 54.77 & $-$26 28 03.6 & $>$27 & 25.44   &   23.16 &  20.78 &  2.76 &  FORS2 &   ...   & 3 \\ 
223   & 11 40 55.99 & $-$26 28 02.9 & 26.13 & 24.30   & $>$23.3 &  21.36 &  2.37 &  FORS2 & 3.455   & 1 \\ 
268   & 11 40 57.02 & $-$26 28 17.5 & $>$27 & 23.02   &   20.57 &  18.35 &  2.64 &  FORS2 &   ...   & 3 \\ 
206   & 11 40 57.87 & $-$26 27 59.3 & $>$27 & 24.52   &   23.51 &  20.28 &  2.40 &  FORS2 &   ...   & 3 \\ 
565   & 11 41 00.10 & $-$26 29 28.1 & $>$27 & $>$26.2 &   22.72 &  20.31 &  2.57 &  FORS2 &   ...   & 4 \\ 
451   & 11 41 00.84 & $-$26 29 04.7 & $>$27 & $>$26.2 & $>$23.3 &  21.26 &  3.11 &  FORS2 &   ...   & 4 \\ 
 90   & 11 41 01.54 & $-$26 27 31.2 & $>$27 & 23.55   &   21.38 &  18.89 &  2.51 &  FORS2 &   ...   & 3 \\ 
728   & 11 41 02.69 & $-$26 30 07.8 & $>$27 & $>$26.2 &   21.03 &  18.89 &  2.56 &  FORS2 &   ...   & 4 \\ 
\hline
... &  11 40 43.85 &  $-$26 31 26.4 & \multicolumn{5}{c}{Ly-$\alpha$ emitter candidate}& FORS2 &   ...   & 3 \\
... &  11 40 50.68 &  $-$26 31 00.0 & \multicolumn{5}{c}{Ly-$\alpha$ emitter candidate}& FORS2 &   ...   & 4\\
\hline
\multicolumn{11}{l}{$^a$Quality flags are the same as in Table~2. Limiting magnitudes are 5$\sigma$.}

\end{tabular}
\end{table*}

\begin{figure*}
\centering
\includegraphics[width=18cm]{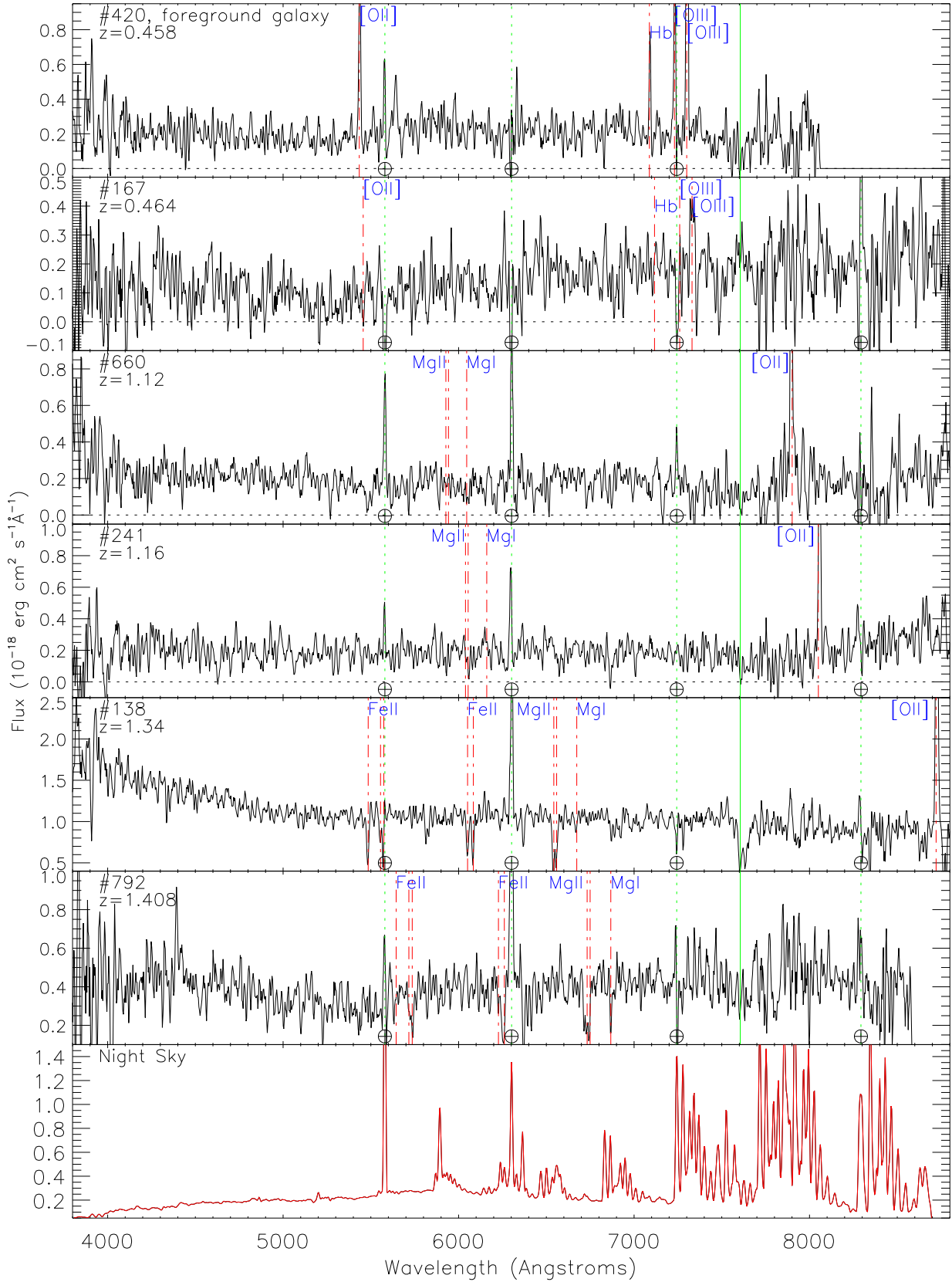} \\
\caption{FORS2 spectra of objects with identified redshifts in the
field of \uss0943. Objects are ordered by redshift from low to high.
The vertical green line indicates the region of atmospheric absorption at
7600 \AA.  The horizontal dashed line indicates zero flux.}
\label{fig:0943spec_opt}
\end{figure*}
\addtocounter{figure}{-1}
\begin{figure*}
\centering
\includegraphics[width=18cm]{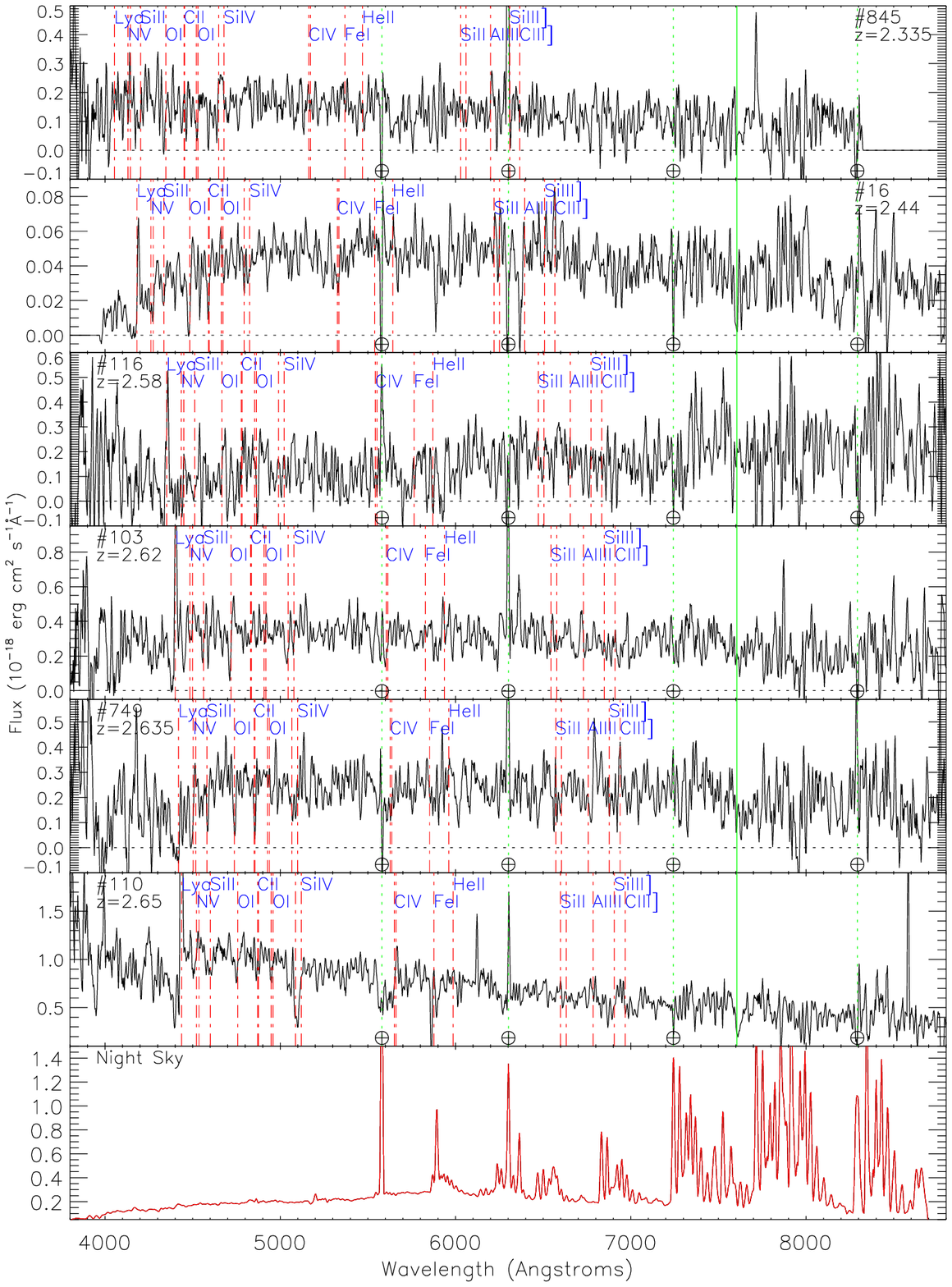} \\
\caption{{\it cont.}}
\end{figure*}
\addtocounter{figure}{-1}
\begin{figure*}
\centering
\includegraphics[width=18cm]{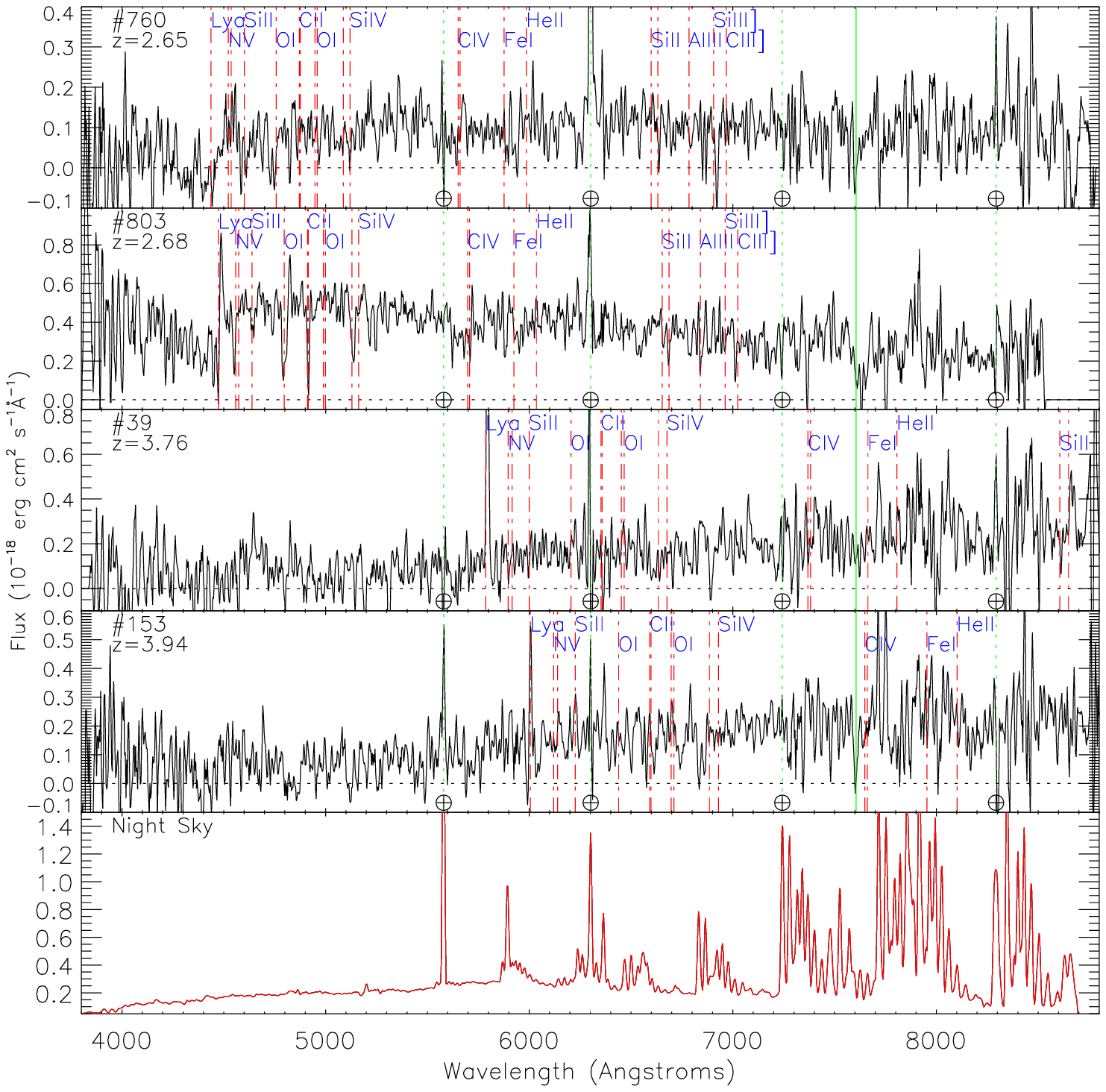} \\
\caption{{\it cont.}}
\end{figure*}

\begin{figure*}
\centering
\includegraphics[angle=0,width=18cm]{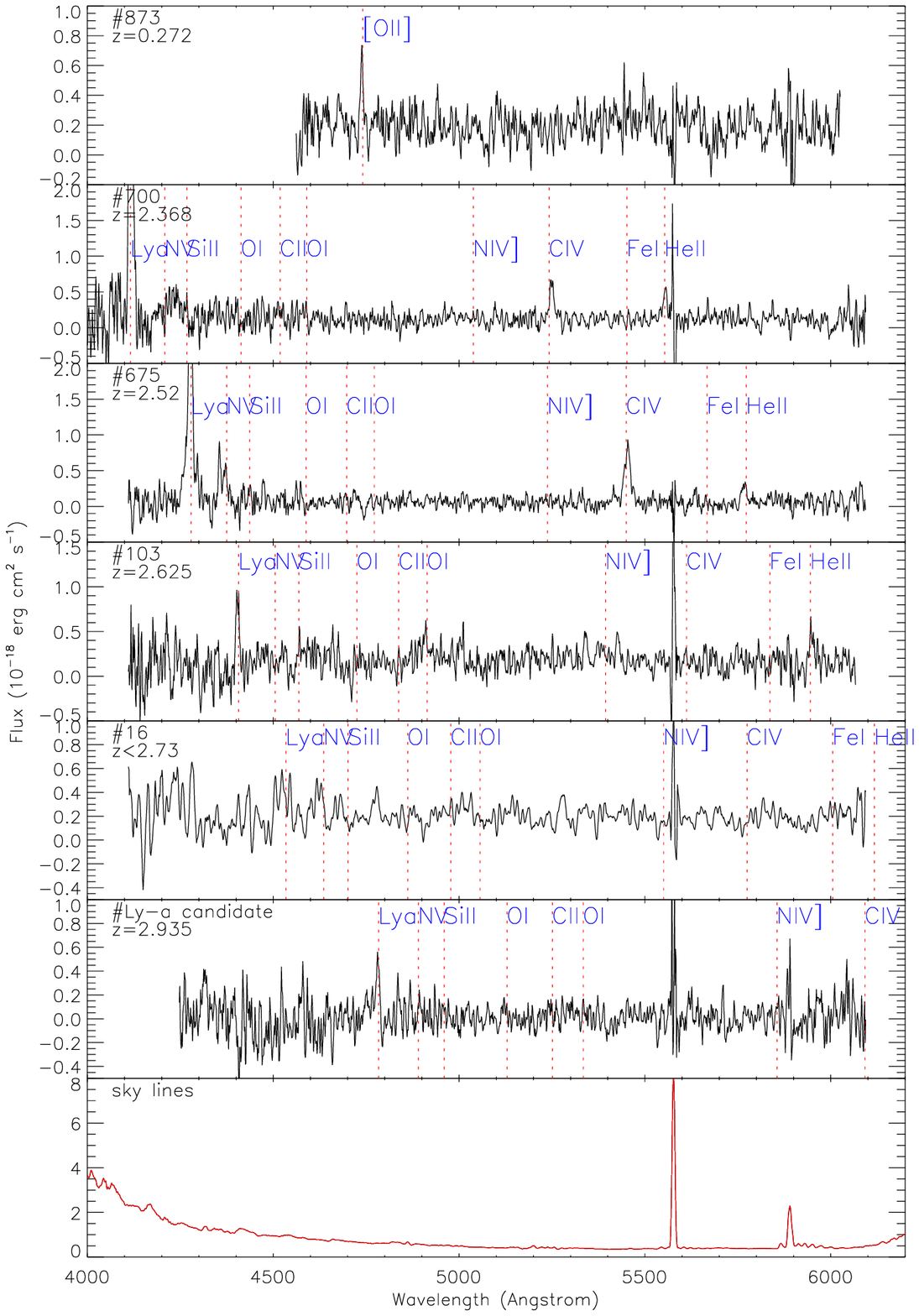}\\
\caption{FOCAS spectra of objects with identified spectra in the
field \uss0943.}
\label{fig:0943spec_foc}
\end{figure*}

\begin{figure*}
\centering
\includegraphics[angle=0,width=18cm]{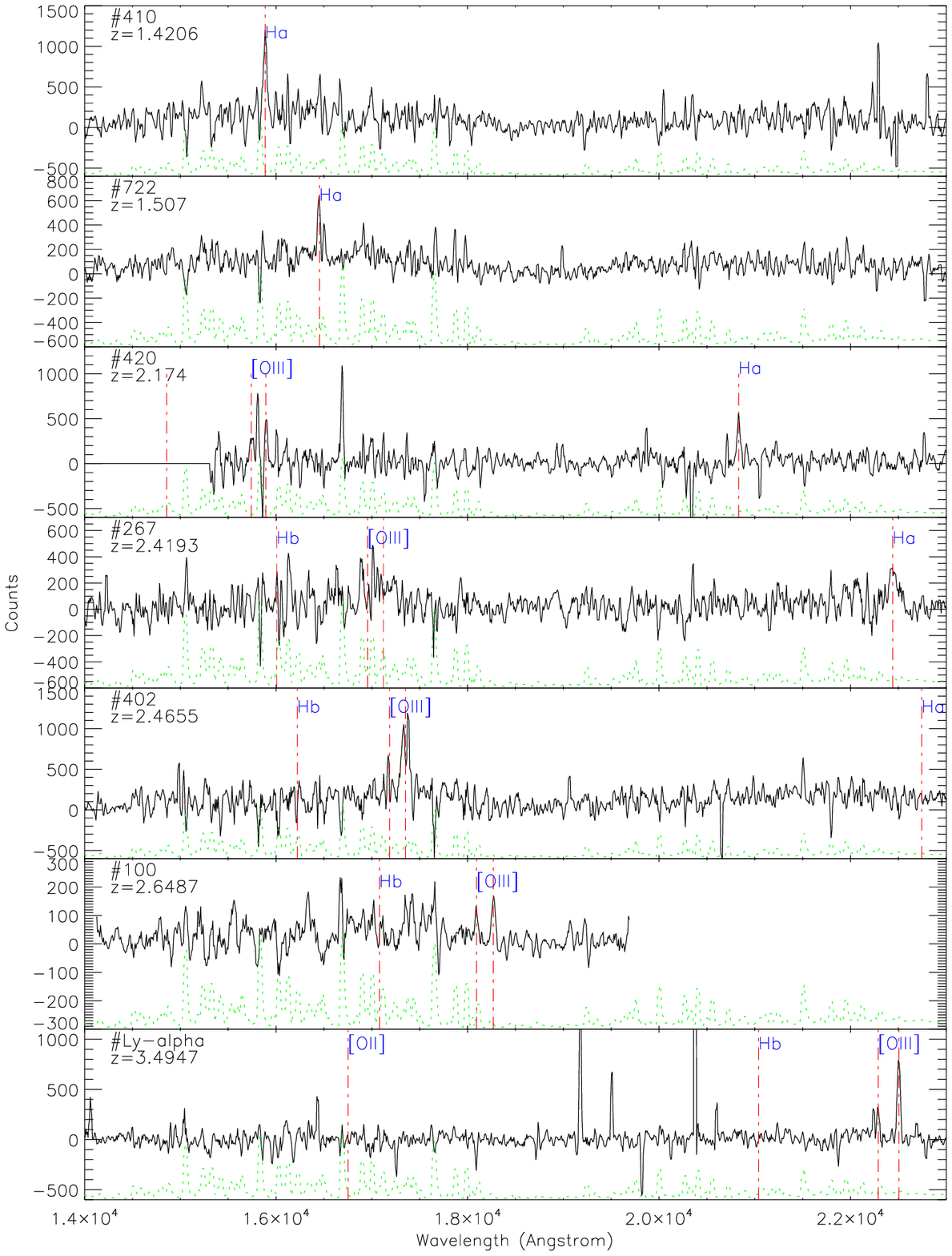}\\
\caption{MOIRCS spectra of objects with identified redshifts in the
field \uss0943.  Sky lines are shown as a dotted green spectrum
beneath each object spectrum. }
\label{fig:0943spec_moircs}
\end{figure*}

Out of 32 objects observed with FORS2 in \uss0943, 16 redshifts have
been identified (Fig.~\ref{fig:0943spec_opt}).  Nineteen objects were
observed with FOCAS, including 17 near-infrared-selected galaxies and
two Ly-$\alpha$ candidates observed as filler objects
(Fig.~\ref{fig:0943spec_foc}).  We obtained five redshifts, only one
of which lies close to the redshift of the radio galaxy, LAE\#381 at
$z=2.935$ (Quality 1) and this is not one of our red galaxy candidate
members but was one of the Lyman-alpha emitter candidates.  The low
success rate with FOCAS is most likely due to the short exposure time
since that observing run was largely weathered out.

We identified a further 8 redshifts with MOIRCS (out of 21 targeted
sources), giving a total of 27 sources in the field of
\uss0943\ with spectroscopic redshifts, out of 57 targeted
(Fig.~\ref{fig:0943spec_moircs}).

\begin{figure}
\centering
\includegraphics[width=\linewidth]{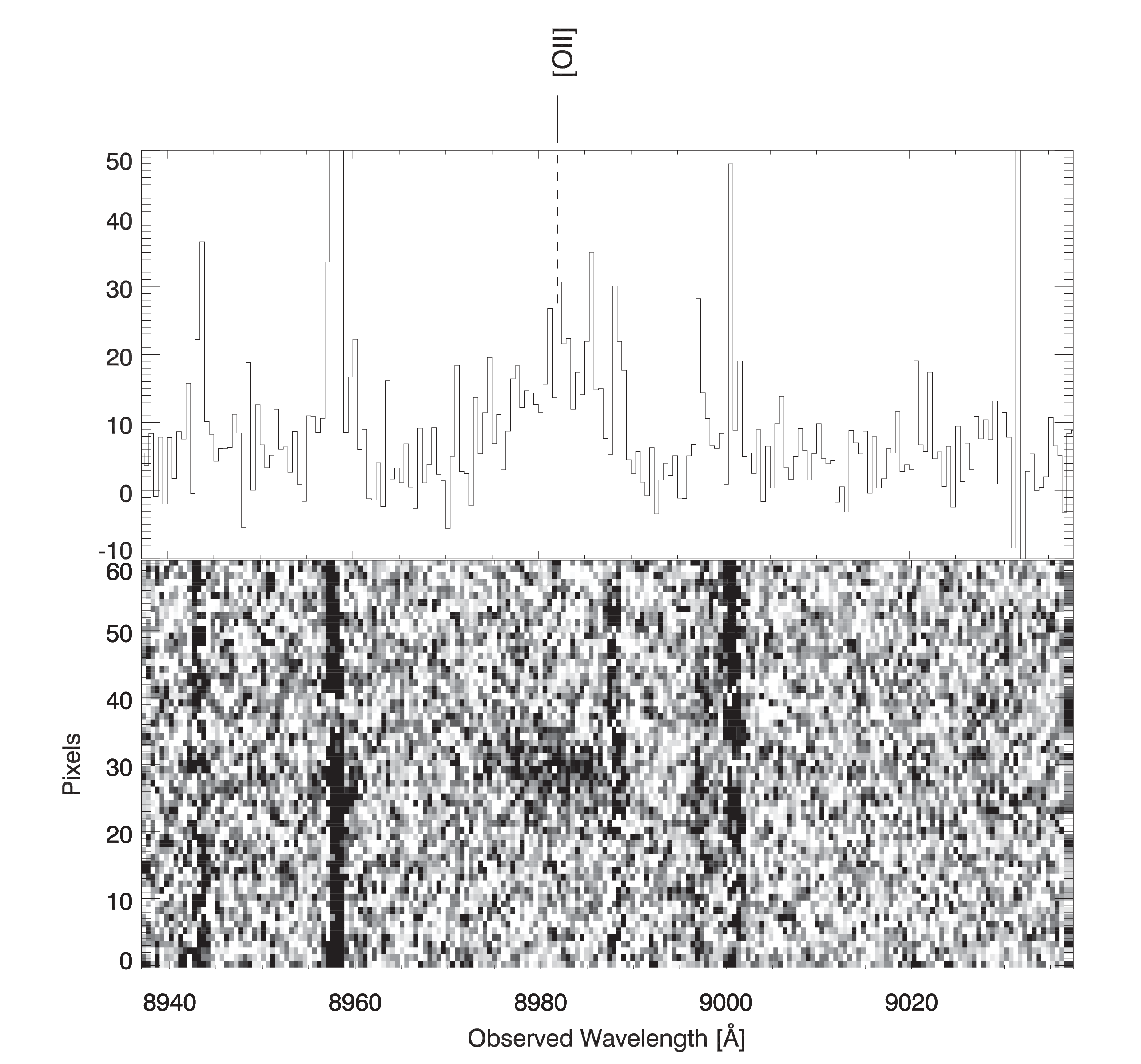}\\
\caption{X-SHOOTER visual--red arm spectrum of object \#792 in the
  field of \uss0943.}
\label{fig:obj792_vis_oii}
\end{figure}

\begin{figure}
\centering
\includegraphics[width=\linewidth]{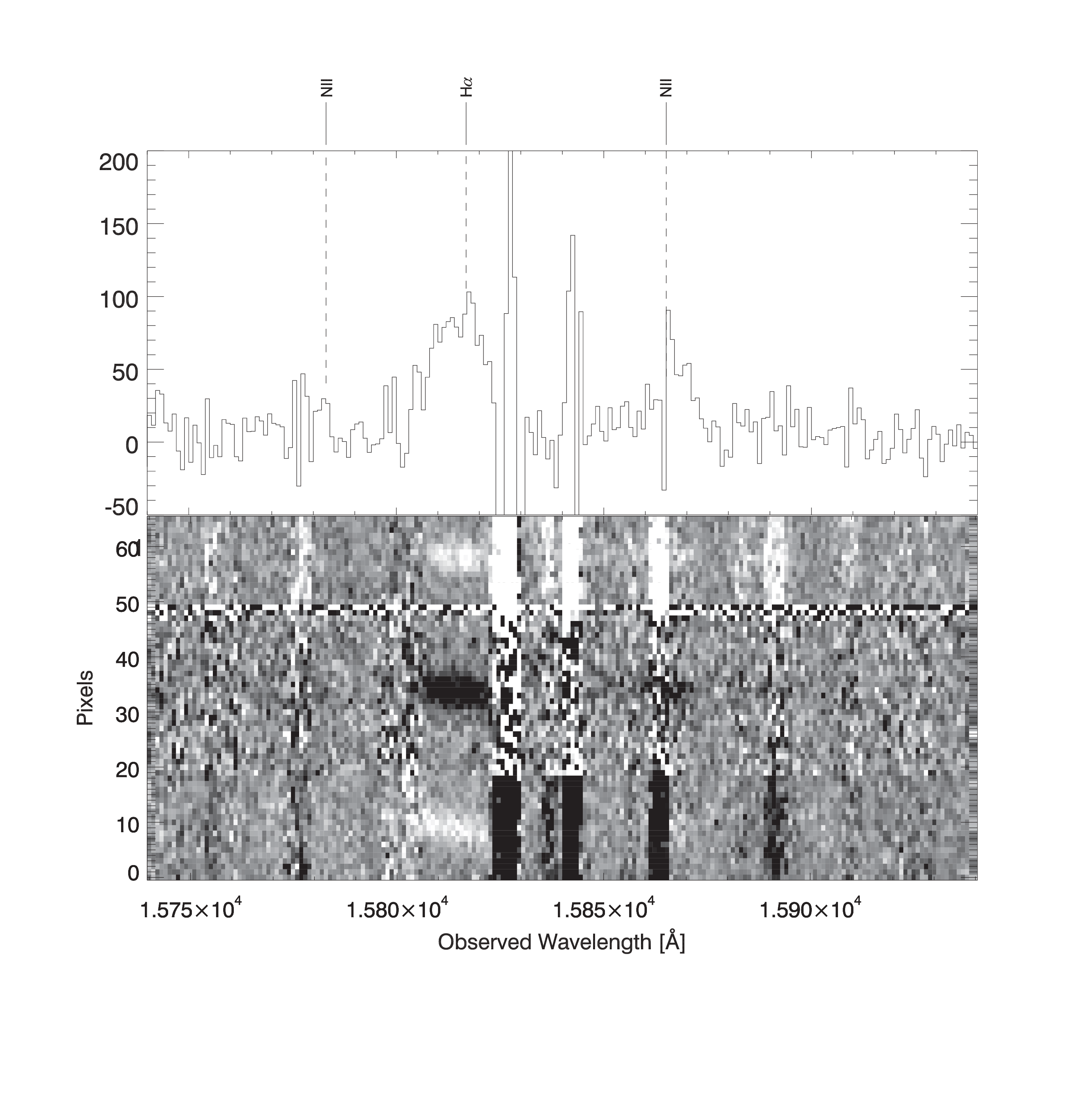}\\
\caption{X-SHOOTER NIR arm spectrum of object \#792 in the field of \uss0943.}
\label{fig:obj792_nir_halpha}
\end{figure}

We also observed one object (\#792) with X-SHOOTER.  The visual-red
arm spectrum (Fig.~\ref{fig:obj792_vis_oii}) shows one line at
8985.7\,\AA. The near-IR arm spectrum
(Fig.~\ref{fig:obj792_nir_halpha}) shows two lines at 15815.7\,\AA\
and 15868.2\,\AA.  We identify those lines as [OII]$\lambda$3727,
H$\alpha$ and [NII]$\lambda$6583 at z=1.410.  This redshift is
consistent with the one determined independently from the spectrum
obtained with FORS.

\begin{figure}[h]
\centering
\includegraphics[width=\linewidth]{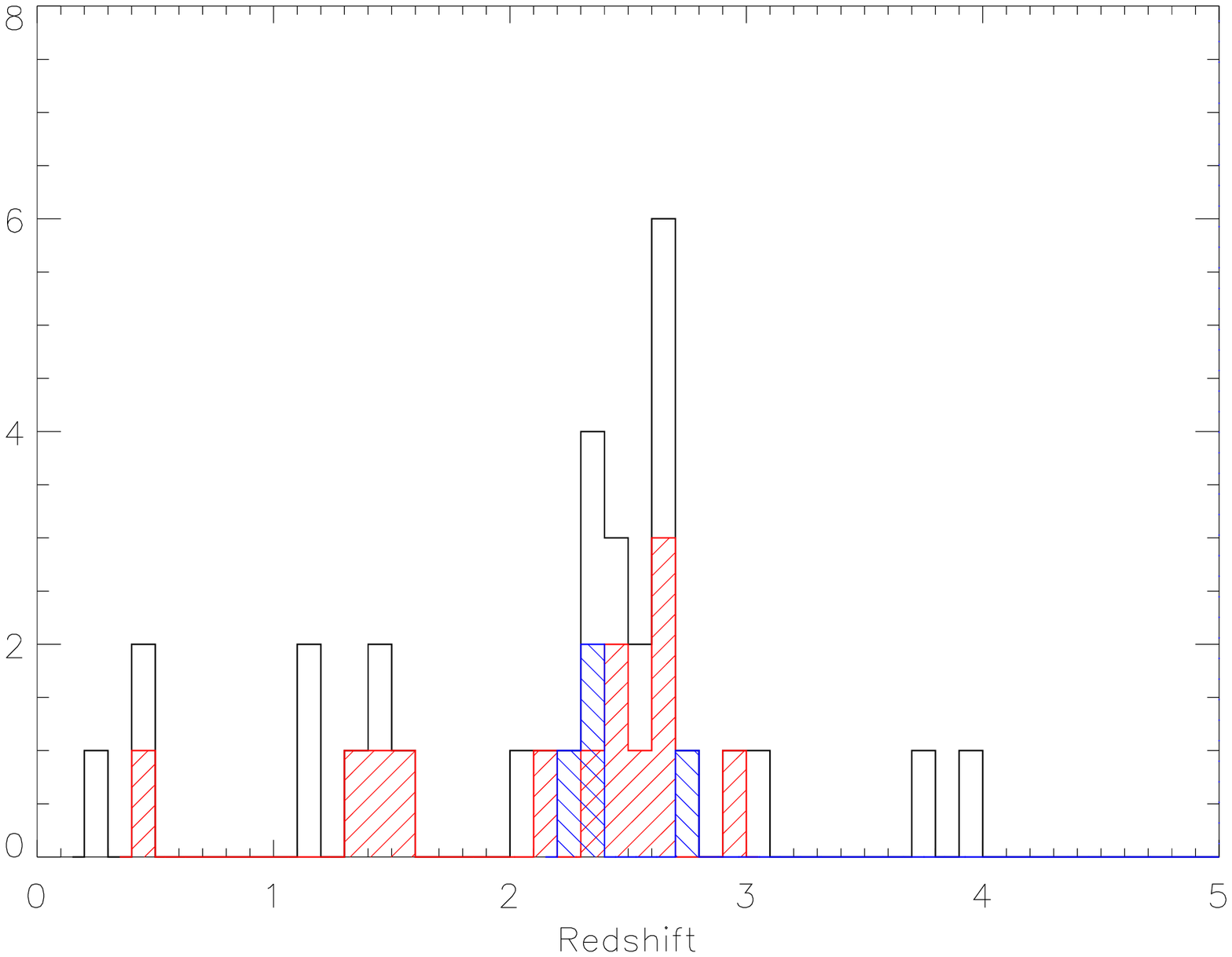}
\caption{Redshift distribution for confirmed sources in the field of
    \uss0943 at $z=2.93$ (redshift indicated by an arrow). The
    proportion of sources in each bin with quality flag 1
    (i.e. reliable redshifts) is shown with red hatching.  The blue
    hatching shows sources with quality 3 (i.e., representing an upper
    limit rather than a confirmed redshift).  No filling (i.e., simply
    the black outline) shows all sources with redshifts (quality flags
    1 and 2).}
\label{fig:zdistrib}
\end{figure}

From the remaining 30 sources, we obtained five redshift upper limits
from the optical spectroscopy, excluding these sources from being
members of the proto-cluster.  The remaining 25 objects cannot be
excluded as either foreground or background objects, and an unknown
fraction of these may even be at the redshift of the radio galaxy.
However, we found an extended foreground structure in this field at
redshift $z \sim 2.6$.  The redshift distribution is shown in
Figure~\ref{fig:zdistrib}.  This foreground structure complicates any
interpretation of the red sequence, or lack thereof, in this
proto-cluster, as a large number of galaxies along the `red sequence'
in \uss0943\ may be members of this intervening structure. However, it
is in itself very interesting as it lies at a high redshift and may
also be a young, forming cluster. We find six galaxies in this
structure, with a mean redshift $<z>=2.65$ and $\Delta z=0.03$.
Several of these sources may contain AGN features (CIV, HeII lines),
though the S/N of our spectra is insufficient to make a clear
statement about this.


For one object, \#420, we obtained two different redshifts from
FORS2 and MOIRCS. Upon closer inspection of the $I$-band and $K-$band
images, we found out that this is actually a pair of galaxies: (i) a
blue one at $z=$0.458 detected in the $I-$band image , and used to
centre our FORS2 spectroscopic slit (Fig.~\ref{fig:0943spec_opt}), and
(ii) a red one at $z=2.174$ detected in the $K-$band image and used to
centre the MOIRCS slit (Fig.~\ref{fig:0943spec_moircs}). These two
galaxies are separated by only 1\farcs1. A hint of the blue galaxy is
seen in the $K-$band image, and vice versa, a hint of the red galaxy
is seen in the $I-$band image, but none of these can be properly
deblended. The red galaxy at $z=2.174$ is the one selected by our
colour criterion.

More than a half (10/18, or 56\%) of the $JHK_s$ selected galaxies
whose redshifts were successfully measured turn out to be actually
located in the redshift range $2.3<z<3.1$.  Of the red sequence
candidates (i.e., rJHK galaxies), 6 out of 10 are within the range
$2.3<z<3.1$.  These numbers are summarised in
Table~\ref{tab:numbersconf}.  The majority of redshift identifications
were based on emission lines, though in several cases the FORS2 data
yielded absorption line identifications where the continuum is strong
enough (e.g., see objects \#749 and \#760 in
Figure~\ref{fig:0943spec_opt}).

\begin{table*}
\caption{Spectroscopy summary for near-IR selected candidates.}
\label{tab:numbersconf}      
\centering                          
\begin{tabular}{r r r r r r}        
\hline\hline
Selection Criterion & \# Candidates & \# Targeted & \# Confirmed Redshifts & \# @ $2.3<z<3.1$ & \# @ $z_{RG}$ \\
\hline 
\multicolumn{6}{c}{\uss0943} \\
rJHK  &  62 & 23 & 10 &  6 & 0 \\
bJHK  &  70 & 15 &  8 &  4 & 1$^a$ \\
Total & 132 & 38 & 18 & 10 & 1$^a$ \\
\hline 
\multicolumn{6}{c}{\pks1138} \\
DRG & 97 & 30 & 15 & 4 & 2 \\
\hline
\end{tabular}\\
$^a$ This is the radio galaxy itself, which is also selected as a bJHK candidate.
\end{table*}

\subsection{\pks1138}

\begin{figure*}
\centering
\includegraphics[angle=90,width=18cm]{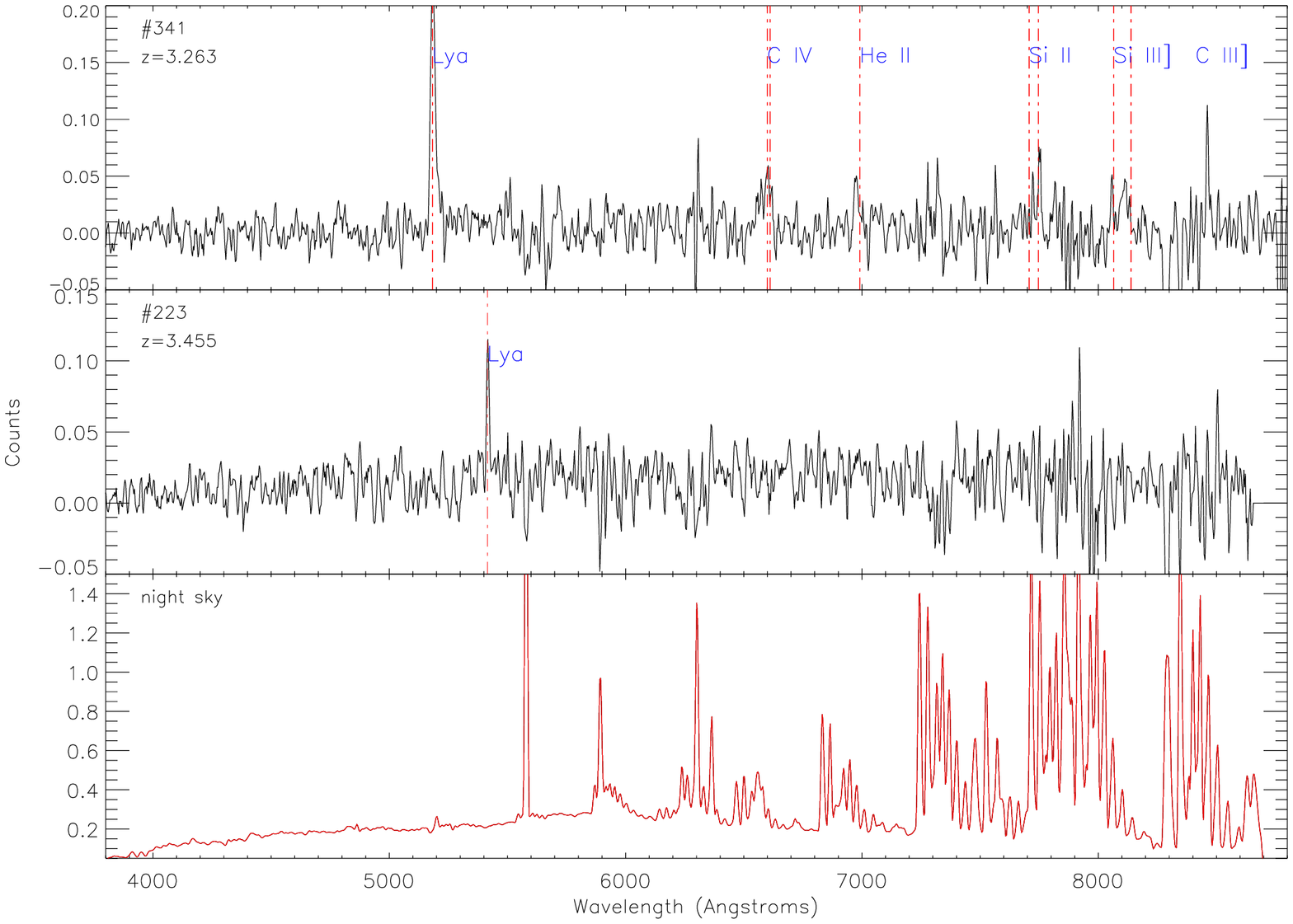}
\caption{FORS2 spectra of background Ly-$\alpha$ emitters identified
in the field of \pks1138.}
\label{fig:1138spec_opt}
\end{figure*}

We observed 27 objects (25 red galaxies and 2 L$\alpha$ emitters) in
the \pks1138 field with optical spectroscopy, but could identify only
two redshifts, both of which are background Ly-$\alpha$ emitters at
$z>3$ (Figure~\ref{fig:1138spec_opt}).  No redshifts were confirmed at
the redshift of the radio galaxy from optical spectroscopy.  Although
the total exposure time is half of that spent on \uss0943, given the
50\% confirmation rate in that field we may naively expect a higher
success rate here given the lower redshift of this radio
galaxy. However, about twice as many sources (25 out of 57) in the
\uss0943 field have $I<24.5$, compared to the \pks1138 field (7 out of
33), so the lower success rate in \pks1138 may be partially due
to the sources being fainter.  We are confident that the astrometry in
both masks is accurate --- confirmed by several stars and bright
galaxies (including the radio galaxy itself) placed on the slitmasks.
The astrometry was internally consistent as the slit positions were
derived from preimaging obtained with FORS2.  However, the efficiency
of the 300V grism drops off quickly at the blue end, and since the
sources observed in this field have such faint magnitudes, it is
likely not possible to detect Lyman breaks at the redshift of the
radio galaxy; we estimate a minimum redshift of $z=2.2$ to detect
Ly-$\alpha$ with FORS2, and the sources were too faint to identify
redshifts from absorption lines.  Furthermore, given the difference in
colour cuts from \uss0943\, which selects intrinsically different
populations (Figure~\ref{fig:targ_sel}), we do not expect to find many
higher redshift interlopers.

\begin{figure}[h]
\centering
\hspace{-0.9cm}\includegraphics[width=1.1\linewidth]{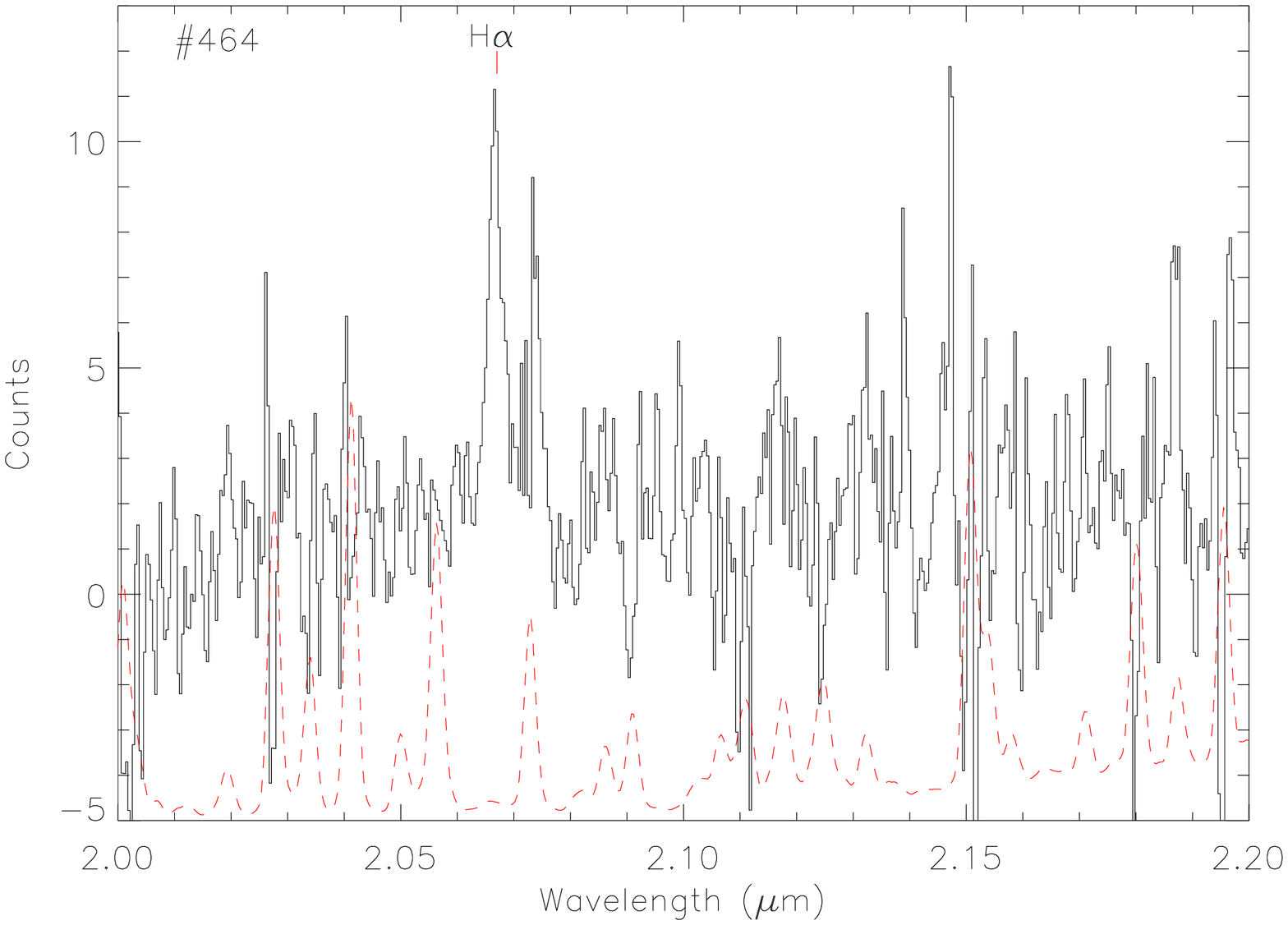}\\
\includegraphics[width=0.95\linewidth]{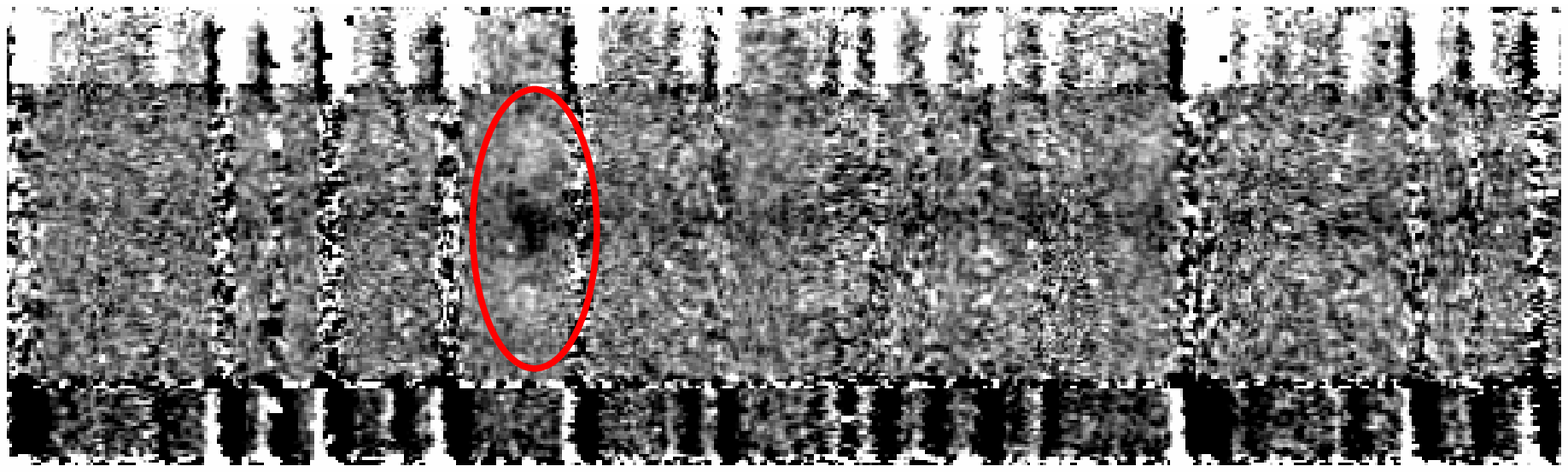}
\caption{(Bottom) 2D and (top) 1D spectra for the object \#464 with
z=2.149. The OH skylines are shown in red underneath the 1D spectrum.
The 2D and 1D spectra are on the same wavelength scale.}
\label{fig:slit03spec-B}
\end{figure}

\begin{figure}[h]
\centering
\hspace{-0.9cm}\includegraphics[width=1.1\linewidth]{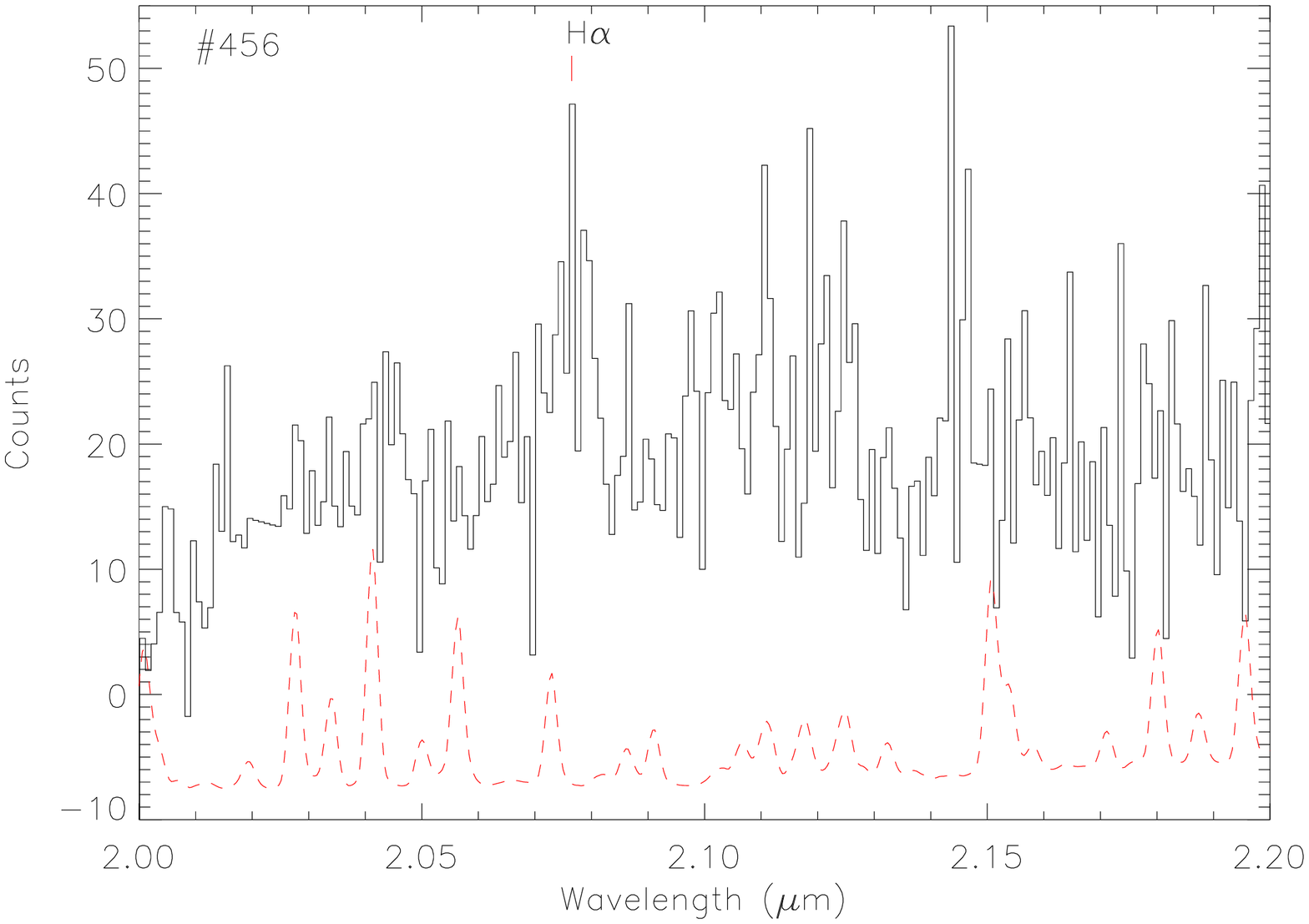}\\
\includegraphics[width=\linewidth]{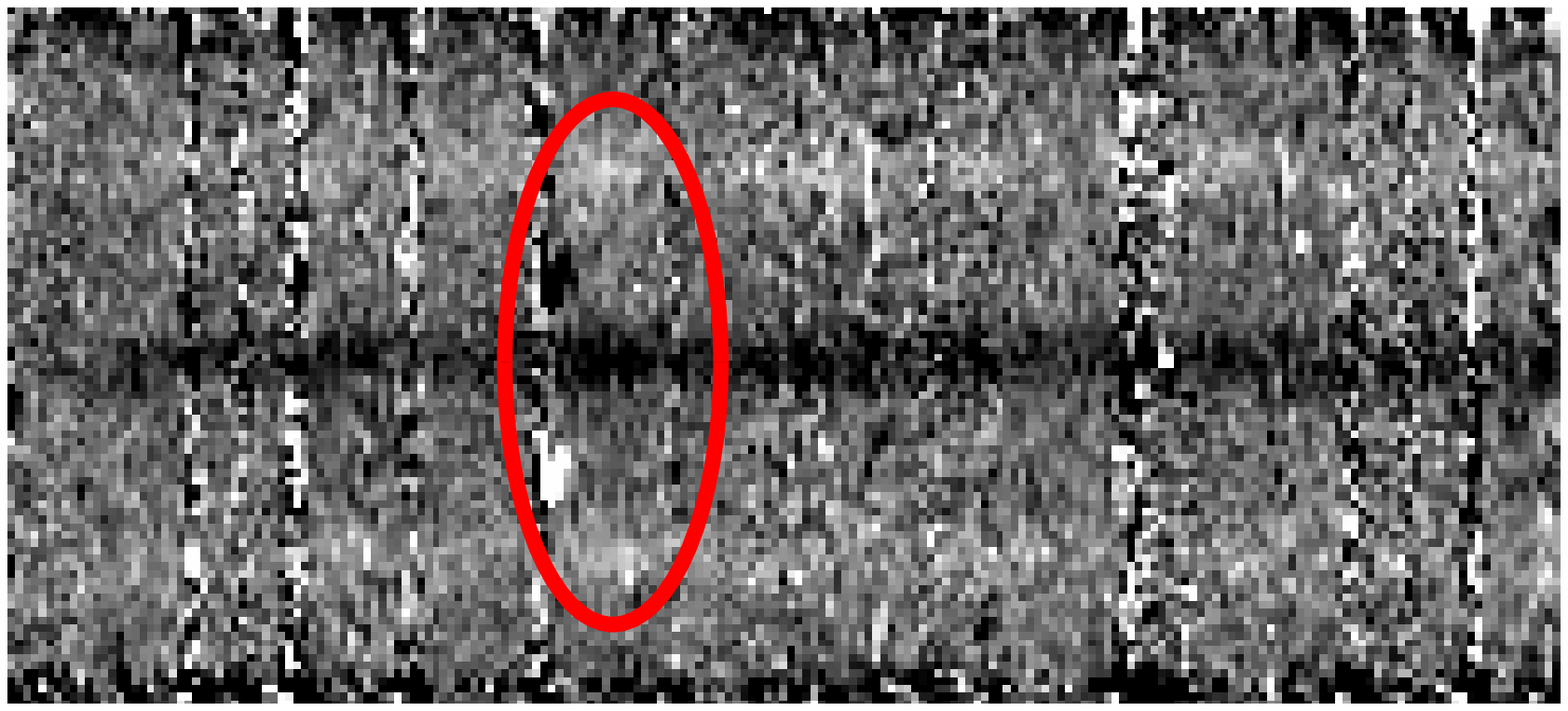}\\
\hspace{0.4cm}\includegraphics[width=0.9\linewidth]{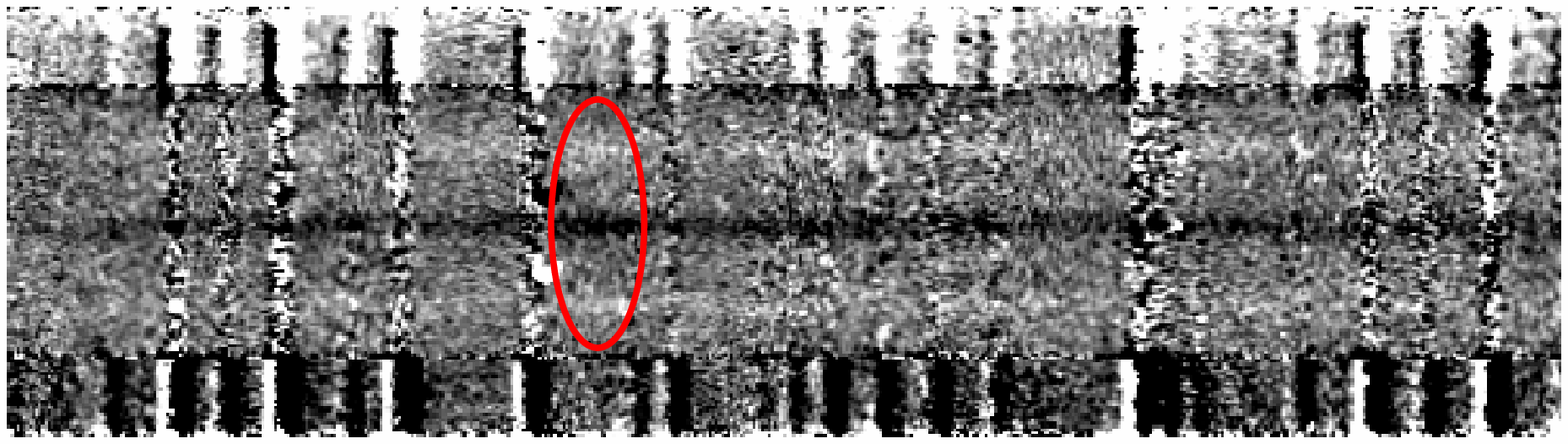}
\caption{Bottom: 2d spectrum  of object \#456 showing H-$\alpha$
detection, Middle: 2d spectrum rebinned in 2x2 pixels to emphasize
the detection and Top: 1d spectrum extracted from the rebinned 2d
spectrum.}
\label{fig:slit02spec}
\end{figure}

From the MOIRCS spectra, we detected two emission line objects (out of
12 targets), which, if identified as H$\alpha$, places the objects at
the same redshift as the radio galaxy.  One of these is a clear
detection, the other is marginal (3$\sigma$).  Both the
two-dimensional and extracted one-dimensional spectra are shown in
Figures~\ref{fig:slit03spec-B} and \ref{fig:slit02spec}. The target
information for these two objects are shown in
Table~\ref{tab:redshifts1138}.

The small number of confirmed redshifts in this field is in fact not
surprising given that detecting interstellar absorption lines and the
Lyman-break in the optical was unsuccessful. As we seem to detect only
a surprisingly low number of star-forming red galaxies in our
colour-selection, the remaining fraction of passively evolving red
sequence galaxies will not have strong emission lines, and hence will
be harder to confirm redshifts of. Our current results show that the
most effective strategy would in fact be to observe much deeper in the
near-IR so that we can detect the 4000\AA\ continuum break feature, if
present \citep[e.g.,][]{kvf+06}.  In fact, there are a total of 13
objects in which we detect continuum in our near-IR spectra, but in
general it is too weak and/or diffuse to identify any features with
the current data.

\section{Properties of the two confirmed red galaxies in \pks1138}

These are the first examples of red galaxies confirmed as members in a
proto-cluster above $z>2$. Here we compare their physical
properties, insofar as possible with the information available, with
the other confirmed proto-cluster members which are generally small,
blue, star-forming, Ly-$\alpha$ and H$\alpha$ emitters.
  
We make use of the multi-wavelength broad-band data to fit spectral
energy distributions (SEDs) and deduce ages, masses and star formation
rates (SFRs) for these two galaxies.  We also calculate the SFRs from
the H$\alpha$ line fluxes in the MOIRCS spectra and from {\it Spitzer}
24\,$\mu$m imaging of the \pks1138 field. Finally, given the projected
spatial location with respect to the RG, and the calculated stellar
masses, we attempt to infer whether or not these galaxies will
eventually merge with \pks1138 .

\subsection{Stellar masses}

Using broad-band $UBRIJHK_s$[3.6][4.5][5.8][8.0] photometry, we
fit the SEDs of the two galaxies confirmed to lie at the redshift of
the \pks1138\ to model templates with two aims.  The first is to test
if the photometric redshifts produced are reliable, which is useful
for future studies.  The second is to derive estimates of the physical
properties of these galaxies --- in particular, their stellar masses.
Tanaka et al. (in prep.) describes the details of the model fitting,
but we briefly outline the procedure here.

The model templates were generated using the updated version of
\citet{bc03} population synthesis code (Charlot \& Bruzual in
prep), which takes into account the effects of thermally pulsating AGB
stars.  We adopt the \citet{sal55} initial mass function (IMF)
and solar and sub-solar metallicities ($Z=0.02$ and $0.008$). We
generated model templates assuming an exponentially decaying SFR with
time scale $\tau$, dust extinction, and age.  We implement effects of
the intergalactic extinction following \citet{furusawa00}, who used
the recipe by \citet{madau95}, as we are exploring the $z>2$ Universe.
We use conventional $\chi^2$ minimizing statistics to fit the models
to the observed data.

Firstly, we generate model templates at various redshifts and perform
the SED fit. Errors of 0.1 magnitudes are added in quadrature to all
bands to ensure that systematic zero point errors do not dominate
the overall error budget.  The resulting photometric redshift of the
object \#456 is $2.25^{+0.06}_{-0.09}$, which is consistent with
the spectroscopic redshift ($z_{\rm spec} = 2.1719$).

We then fix the redshift of the templates at the spectroscopic
redshift and fit the SED again.  We impose the logical constraint that
the model galaxies must be younger than the age of the Universe.  The
derived properties of the galaxy \#456 are an age of
$1.6^{+1.1}_{-0.7}$ Gyr, an e-folding time scale
$\tau=0.1^{+0.4}_{-0.1}$ Gyr, and dust extinction of
$\tau_V=0.4^{+1.4}_{-0.4}$, where $\tau_V$ is the optical depth in the
$V$-band.  It is a relatively old galaxy with a modest amount of
dust, although the error on the extinction is large.  The
apparent red color of the galaxy is probably due to its old stellar
populations.  We find a high photometric stellar mass of
$2.8^{+1.5}_{-1.0}\times 10^{11}\ \rm M_\odot$ for this galaxy.  The
short time scale derived ($\tau$ of only 0.1 Gyr) suggests that
the galaxy formed in an intense burst of star formation. The low
current star formation rate of $0.0^{+1.0}_{-0.0} \rm~M_\odot\
yr^{-1}$ derived from the SED fit indeed confirms that the high star
formation phase has ended and the galaxy is now in a quiescent phase.

For the galaxy \#464, the best fit photometric redshift is $z_{\rm
phot}=2.05^{+0.26}_{-0.13}$, consistent with the spectroscopic
redshift ($z_{\rm spec} = 2.149$).  The best fit at the spectroscopic
redshift yields an age of $2.4^{+0.4}_{-1.0}$ Gyr, an e-folding time
$\tau=0.2^{+0.1}_{-0.2}$ Gyr, dust extinction of
$\tau_V=0.7^{+0.7}_{-0.6}$, a stellar mass of $5.1^{+1.5}_{-2.0}\times
10^{11}\ \rm M_\odot$ and a SFR of $0.0^{+1.0}_{-0.0}\ \rm M_\odot\
yr^{-1}$.

\begin{figure}
\centering
\includegraphics[width=\linewidth]{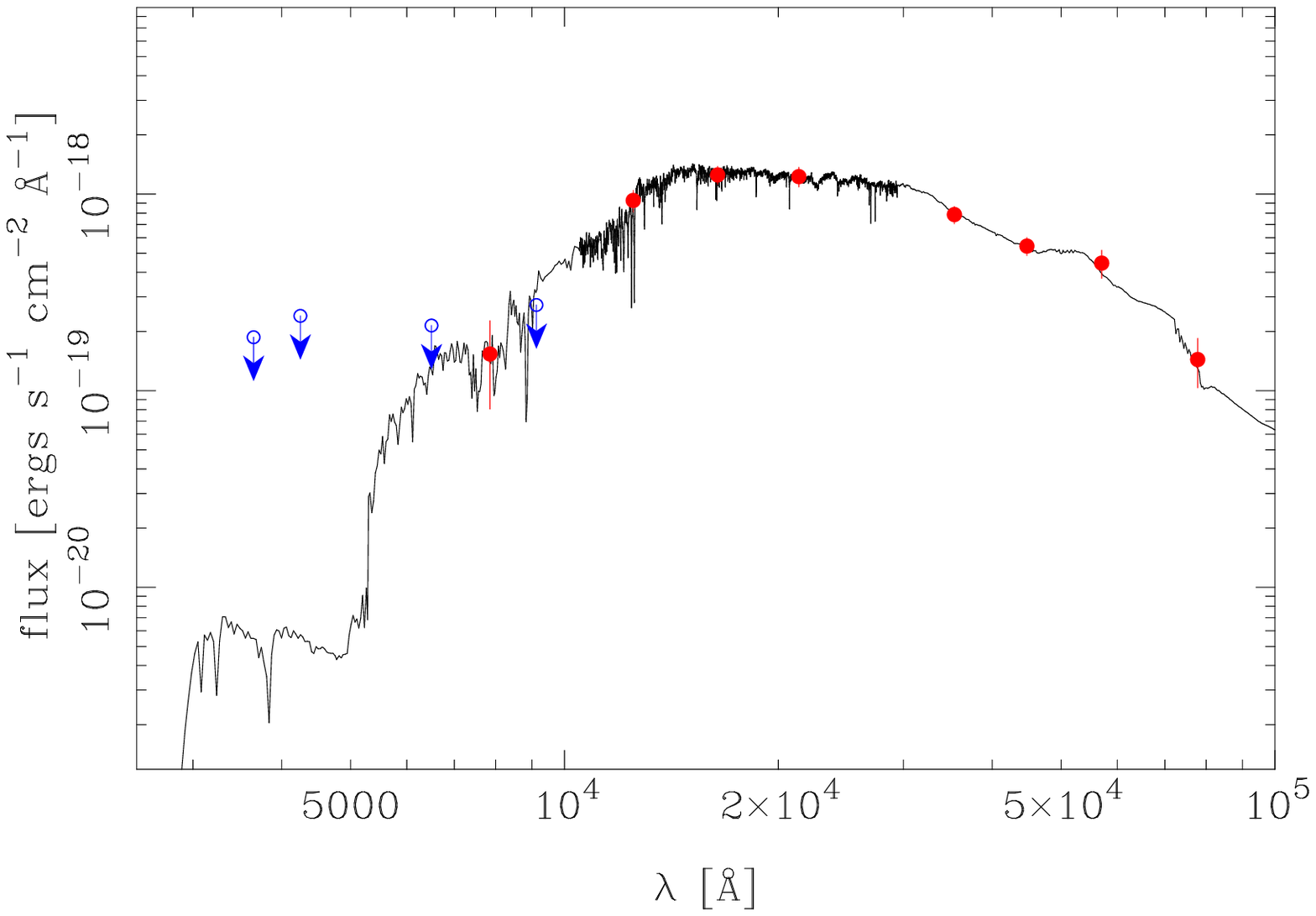}
\caption{SED fit for object \#456, with the fainter H$\alpha$ emission line detection.}
\label{fig:sed456}
\end{figure}

\begin{figure}
\centering
\includegraphics[width=\linewidth]{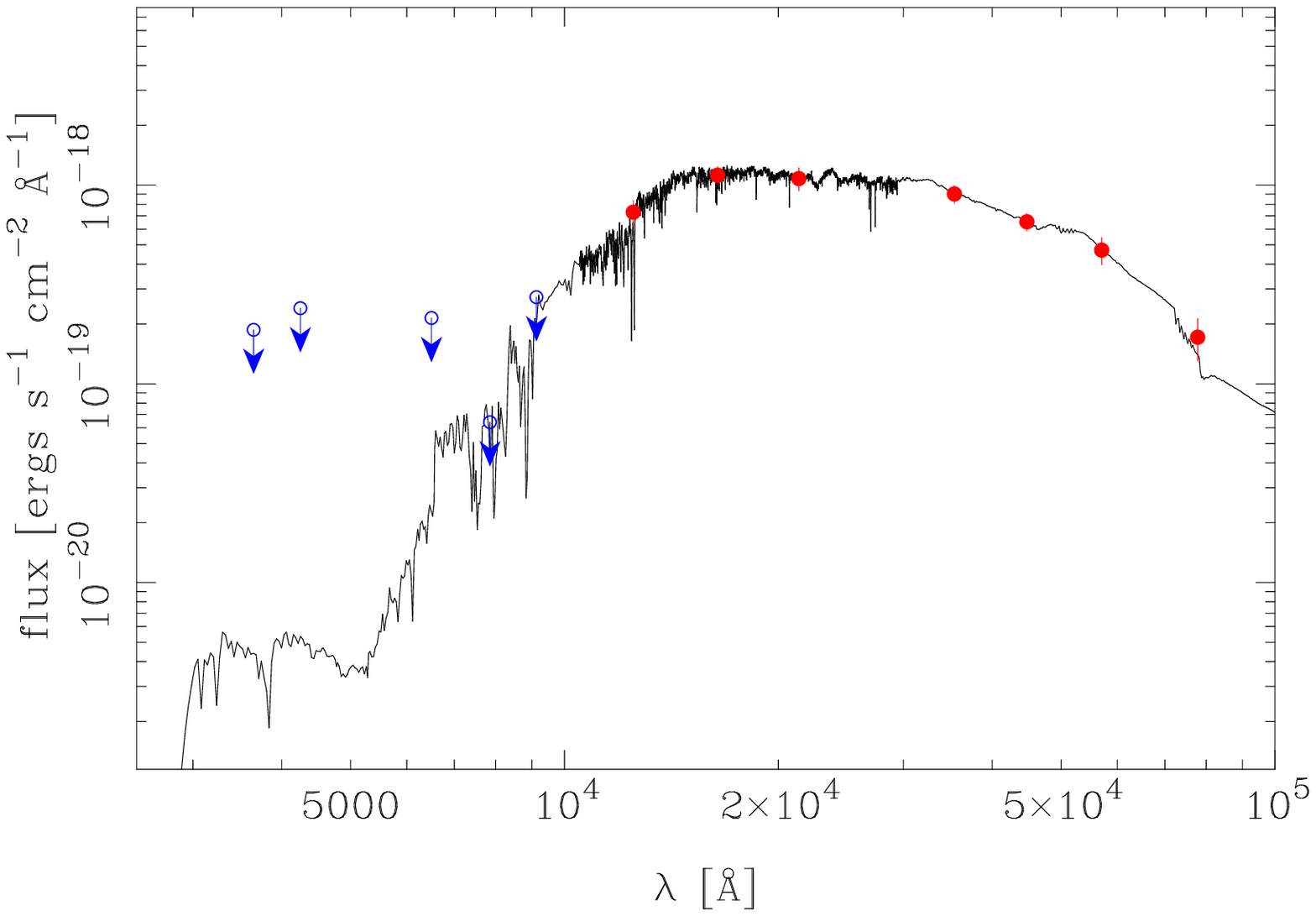}
\caption{SED fit for object \#464, with the brighter H$\alpha$ emission line detection.}
\label{fig:sed464}
\end{figure}

The best fitting SEDs for each galaxy are shown in
Figures~\ref{fig:sed464}~and~\ref{fig:sed456}.  The broad-band
magnitudes of the two galaxies are summarized in Table
\ref{tab:mags1138}.  A more extensive analysis of the properties of
galaxies in the \pks1138\ field will be presented in Tanaka et al. (in
prep.).

\begin{table*}
\caption{Total magnitudes of the two H$\alpha$ detected galaxies.  In
this table, the magnitudes are on the AB system for the ease of
fitting SEDs.  Magnitude limits are $3\sigma$ limits.}
\label{tab:mags1138}
\centering
\tiny\hspace*{-1.cm}
\begin{tabular}{@{}ccccccccccccc@{}}
\hline\hline \\
ID  &  $U$     & $B$     & $R$     & $I$            & $Z$     & $J$            & $H$            & $K_s$          & $3.6\mu m$     & $4.5\mu m$     & $5.8\mu m$     & $8.0\mu m$\\
456 &  $>26.6$ & $>26.0$ & $>25.2$ & $25.15\pm0.41$ & $>24.2$ & $22.19\pm0.08$ & $21.27\pm0.04$ & $20.72\pm0.06$ & $20.11\pm0.03$ & $20.00\pm0.04$ & $19.69\pm0.13$ & $20.24\pm0.25$\\
464 &  $>26.6$ & $>26.0$ & $>25.2$ & $>26.1$        & $>24.2$ & $22.45\pm0.11$ & $21.39\pm0.04$ & $20.86\pm0.09$ & $19.96\pm0.03$ & $19.80\pm0.03$ & $19.63\pm0.12$ & $20.05\pm0.21$\\
\hline \\
\end{tabular}
\end{table*}

Several earlier papers presented measurements of different
  classes of objects in the field of \pks1138. \citet{kprm04}
converted K-band magnitudes to masses, for the Ly-$\alpha$ and
H$\alpha$ emitters, estimating masses of $0.3-3\times10^{10} \rm
M_{\odot}$ for the Ly-$\alpha$ emitters (except for one object
  with a mass of $7.5\times10^{10} \rm M_{\odot}$) and
  $0.3-11\times10^{10} \rm M_{\odot}$ for the H$\alpha$ emitters, the
  latter on average being clearly more massive than the Ly-$\alpha$
  emitters. Furthermore \citet{hok+09} find stellar masses
  between $4\times10^8$ and $3\times10^{10}$ for candidate Ly-$\alpha$
  emitting companions within 150~kpc of the central RG. Our red
  galaxies are thus an order of magnitude more massive than the most
  massive Ly-$\alpha$ emitters and 2--3 times more massive than the
  most massive H$\alpha$ emitters discovered through NB imaging
  searches in this field.

\subsection{Star formation rates}

We now use the line fluxes for the two H$\alpha$ detections to infer a
lower limit to the instantaneous SFR.  
We obtained a rough flux calibration of the spectra by scaling the
total flux in the spectrum to the observed $K_S$magnitude of the two
objects, and assuming a flat spectral throughput in the $K_S$-band. As
the observed lines are near the centre of the $K_S$-band, this
approximation does not significantly effect the derived H$\alpha$
flux, and we assume a calibration uncertainty of $\approx 20\%$.  Both
line detections lie on the edge of weak telluric absorption features
between 20400-20800 \AA\ and also very close to sky lines (especially
\#458), increasing the measurement uncertainties.  The H$\alpha$ flux
of object \#464 thus derived is $1.3\pm0.3 \times 10^{-16}$ erg
s$^{-1}$ cm$^{-2}$ and the fainter object, \#456, has an H$\alpha$
flux of $6.2\pm2 \times 10^{-17}$ erg s$^{-1}$ cm$^{-2}$.  The
measured line flux in \#464 is fully consistent with the published
F(H$\alpha$)=$1.35\times 10^{-16}$ erg s$^{-1}$ cm$^{-2}$ derived from
narrow-band imaging \citep[Object 229 of ][]{kprm04}. The previously
published near-IR spectroscopy of \citet{kpo+04} found
F(H$\alpha$)=7.1$\pm1.9\times 10^{-17}$ erg s$^{-1}$ cm$^{-2}$, which
is somewhat lower than our flux, suggesting possible slit losses. In
the same narrow-band image, object \#458 does not show any excess flux
compared to the full $K-$band image, suggesting that our line flux may
be overestimated due to incomplete skyline subtraction.

Using the \citet{ken98} relation, $$ {\rm SFR}\, ({\rm M}_{\odot}\,
{\rm yr}^{-1}) = 7.9 \times 10^{-42}\ L({\rm H}\alpha)\, {\rm erg}\,
{\rm s}^{-1},$$ which assumes solar metallicities and a \cite{sal55}
IMF, this leads to SFRs of 35$\pm$8 and 17$\pm$6 M$_{\odot}$ yr$^{-1}$,
respectively.


The SFRs derived from H$\alpha$ are higher than those from the SED
fits (Figs \ref{fig:sed456} and \ref{fig:sed464}), which give SFR up
to $4 \rm M_\odot\ yr^{-1}$ at $2 \sigma$, i.e. a factor of 5--10
lower.  An additional independent estimate for the SFR can be obtained
from the MIPS 24$\mu$m imaging of the field \citep{ssd+07}. Object
\#464 is detected at S(24$\mu$m)=470$\pm$30\,$\mu$Jy, while object
\#458 remains undetected at at the 2$\sigma$$<$$60\,\mu$Jy level. We
first convert the 24$\mu$m flux to the total IR flux using the
relation of \citet{red06}, and converted the latter to a SFR using the
formula of \citet{ken98}. The derived SFR are 34 and $<$4\,M$_{\odot}$
yr$^{-1}$ for objects \#464 and \#458, respectively. For \#464, this
is fully consistent with the the SFR derived from H$\alpha$, but for
\#458, the value derived from H$\alpha$ seems strongly
over-estimated. This may be either be due to the low S/N and
incomplete skyline removal in our near-IR spectroscopy, or it may
indicate that the H$\alpha$ may have a contribution by an AGN. To
confirm the latter hypothesis, we need to obtain other emission lines
such [OIII], but the possible AGN contribution may not be surprising
given the higher fraction of AGNs observed in proto-clusters
\citep[e.g.][]{gse+09}.

In summary, we find that object \#458 most likely belongs to the class
of passively evolving galaxies, while \#464 belongs to the class
of dusty star-forming red galaxies.

\subsection{Merging timescales }

\begin{figure}[h]
\centering
\includegraphics[width=\linewidth]{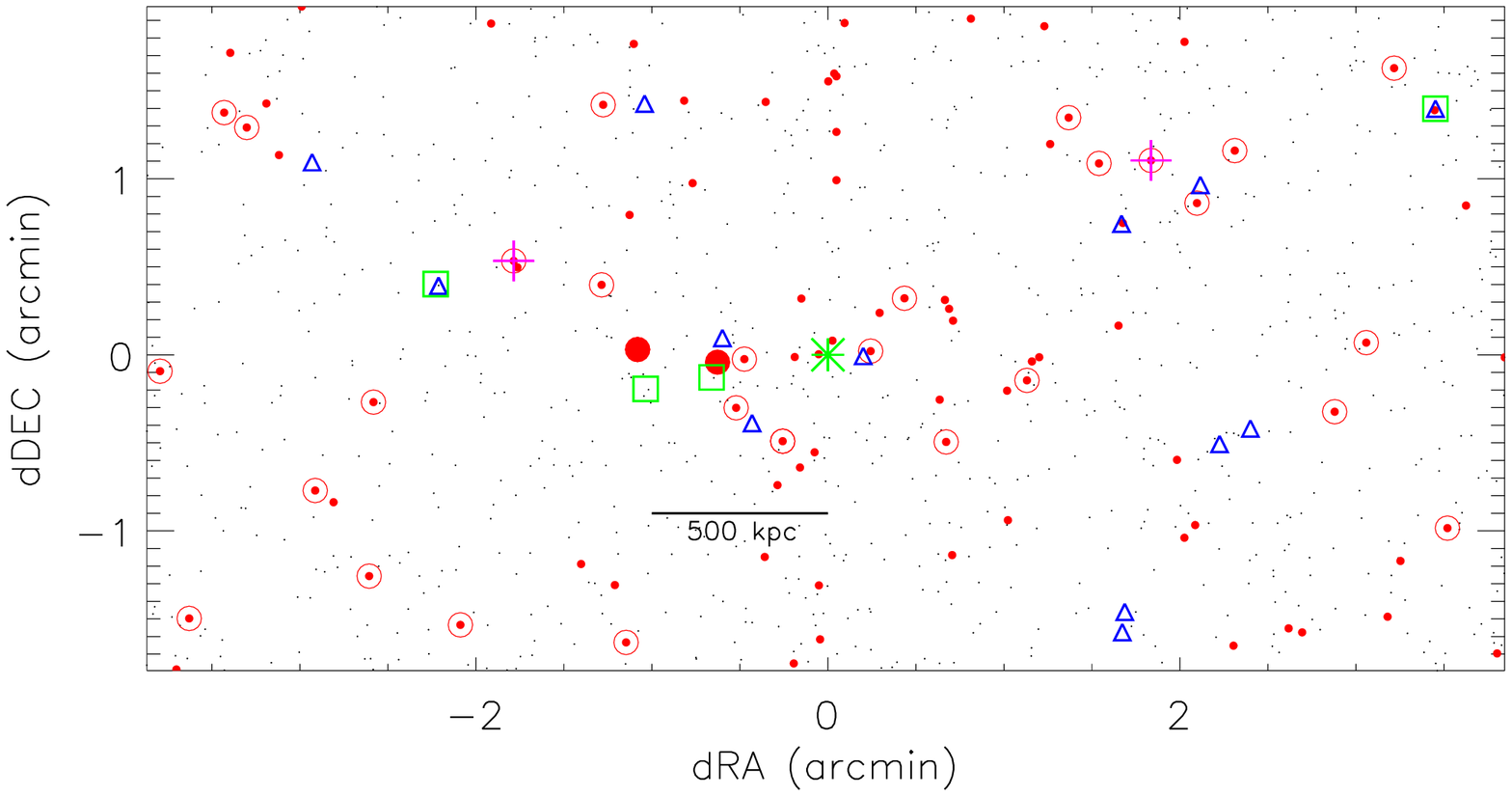}
\caption{Relative spatial distribution of the two confirmed massive
galaxies in \pks1138.  The central radio galaxy is marked by a large
green asterisk.  The red galaxy candidates are marked as red points;
those targeted spectroscopically are marked with circles. The two
filled red circles are the galaxies confirmed to lie at the redshift
of the radio galaxy.  Confirmed Lyman-$\alpha$ emitters are marked
as blue triangles \citep{pkr+00}, and confirmed AGN are marked as
the smaller green squares \citep{ckv+05,pkc+02}. Objects which have
been excluded as background objects are marked with a plus sign.
}
\label{fig:radec1138}
\end{figure}

The spatial locations of the two red galaxies identified at the
redshift of \pks1138\ are shown in Figure~\ref{fig:radec1138}.  They
both lie along the East-West axis, where most of the confirmed AGN in
this field were found \citep{ckv+05, pkc+02} and where the brighter
DRGs in \citet{kkk+07} also trace a filament.  The two galaxies are
relatively central, being located within 1\arcmin\ (500 kpc,
  physical) of the radio galaxy.  This is significant given that no
obvious density gradient has been observed for the Lyman-$\alpha$
emitters \citep[e.g.,][]{pkr+00, krp+03}, yet \citet{kprm04} argue
there is an indication that the H$\alpha$ emitters and extremely
red objects have a higher density near the RG than further out. Our
new results and these former results are consistent with the picture
that accelerated evolution takes place in high density environments.

To properly assess the possibility of a future merger between the RG
and one or both of the red companion galaxies, numerical simulations
are needed. However, we can at least gain an idea of the relative
timescales involved as follows.  The crossing time can be approximated
by $t_{\rm cross}=(R^3/GM_{\rm total})^{1/2}$ \citep{bt08} where R is
the distance between the two galaxies and M is the total mass (dark
matter + stellar) of the central galaxy, approximately $10^{13} \rm
M_{\odot}$ in this case \citep{hok+09}. For the closer galaxy to the
RG, \#464, the distance is $R\sim300$~kpc (physical) and the crossing
time is therefore roughly 0.7~Gyr. The second galaxy has a crossing
time between 2--3 times longer, let us say approximately 2~Gyr. For a
major merger (unless it is a high speed encounter in which case the
galaxies would pass through each other with little disruption) the
galaxies should merge into a single system within a few crossing times
\citep{bt08}. There is therefore enough time for the two red galaxies
to merge with the central RG before redshift $z=0$, increasing its
stellar mass of $10^{12} \rm M_{\odot}$ \citep{hok+09} by a factor of
at least 50\% .

\section{Conclusions}

We observed a total of 57 near-infrared selected galaxies in \uss0943\
and 33 in \pks1138\ with a mix of optical and near-infrared
multi-object spectroscopy to attempt the confirmation of massive
galaxies in the vicinity of the central radio galaxy in these
candidate forming proto-clusters.  We have determined that the $JHK_s$
selection criteria presented in \citet{kkt+06} select mostly
$2.3<z<3.1$ galaxies, thus confirming the validity of this colour
selection technique (10 out of 18 confirmed redshifts of
$JHK_s$-selected galaxies are in this redshift range).  Of the 57
sources in the field of the radio galaxy \uss0943 we identify 27
spectroscopic redshifts and for the remaining 30 sources we obtained
five upper limits on the redshifts, excluding them to be part of the
proto-cluster.  The remaining 25 objects could not be excluded as
either foreground or background. An unknown fraction of these may be
at the redshift of the radio galaxy.
However, we also pinpoint a foreground (but still quite distant) large
scale structure at redshift $z\approx 2.65$ in this field, so it is likely
that some of the remaining galaxies are also associated with this
structure.

Of the 33 galaxies observed in \pks1138\ we exclude two background
objects and confirm two objects at the redshift of the radio galaxy
(all on the basis of emission lines).  We cannot exclude the remaining
29 objects as being potential proto-cluster members --- we require
deep near-infrared spectroscopy to locate the 4000\AA\ break. On
the basis of SED fits, and the SFR derived from both the H$\alpha$
line flux and the 24\,$\mu$m flux, we deduce that one of the two red
galaxies confirmed to be at the redshift of the proto-cluster belongs
to the class of dusty star-forming (but still massive) red galaxies,
while the other is evolved and massive, and formed rapidly in intense
bursts of star formation.  This represents the first red galaxies to
be confirmed as a member of a proto-cluster above $z>2$. The only
other galaxies of the same class are evolved galaxies found in an
overdensity at $z=1.6$ in the Galaxy Mass Assembly ultra-deep
Spectroscopic Survey \citep{kcz+09}, and red massive galaxies in the
Multi-wavelength Survey by Yale-Chile (MUSYC) sample at $z\sim2.3$
\citep[e.g.~][]{kvf+06}. Since their phase of high star formation has
ended (the current SFR is of order 1~M$_{\odot}$ yr$^{-1}$), we only
just detect them through their emission line fluxes. However, this
shows that even in these galaxies where we do not expect strong
emission lines, it is still sometimes possible to derive redshifts
from the H$\alpha$ line.  It is highly likely that there are more
galaxies on the red sequence in this field which have completely
ceased star formation and are thus very difficult to identify
spectroscopically. The next stage of our spectroscopic campaign is to
obtain deep near-infrared spectroscopy in the $J$- and $H$-bands to
search for objects with 4000\AA\ continuum breaks.

Given the inferred SFRs and stellar masses of the two confirmed
$z\sim2.15$ galaxies, they have to have formed quite early, which fits
into the down-sizing scenario \citep{cshc96}. Although this conclusion
is drawn from limited numbers, the low fraction of sources with
H$\alpha$ emission lines in our sample suggests that most of the
sources probably don't have much on-going star formation (SFR$<0.5 \rm
~M_{\odot} ~yr^{-1}$) and so their formation epoch might be quite
high. This would be contrary to the idea of the giant elliptical
assembly epoch being $z\sim2-3$ \citep[e.g.,][]{vfk+08, kvv+08}, but
is consistent with results from SED age fitting of stellar
populations, which point to $z_{form}>3$ (e.g., Eisenhardt et
al. 2008\nocite{ebg+08}).  The discrepancy may be resolved if
subclumps form early and merge without inducing much star formation
(i.e. dry merging). Alternatively, the fact that these galaxies
  are located in over--densities may predispose us to structures that
  formed early.

Finally, the proximity of these two massive galaxies to the RG implies
that they will have an important impact on its future evolution. Given
the crossing times, compared with the time remaining until $z=0$, it
is plausible that one or both of the galaxies may eventually merge
with the RG.

\begin{acknowledgements}
We thank the referee G. Zamorani for a very constructive referee
  report, which has substantially improved this paper.  This work was
  financially supported in part by the Grant-in-Aid for Scientific
  Research (No.s 18684004 and 21340045) by the Japanese Ministry of
  Education, Culture, Sports and Science. CL is supported by NASA
  grant NNX08AW14H through their Graduate Student Researcher Program
  (GSRP).  We thank Dr. Bruzual and Dr. Charlot for kindly
  providing us with their latest population synthesis code.  MD
  thanks Andy Bunker and Rob Sharp for useful discussions on
  manipulating MOIRCS data.  The work of DS was carried out at Jet
  Propulsion Laboratory, California Institute of Technology, under a
  contract with NASA. JK acknowledges financial support from DFG grant
  SFB 439.  The authors wish to respectfully acknowledge the
  significant cultural role and reverence that the summit of Mauna Kea
  has always had within the indigenous Hawaiian community. We are
  fortunate to have the opportunity to conduct scientific observations
  from this mountain.

\end{acknowledgements}

\bibliography{myrefs}
\bibliographystyle{aa}

\end{document}